arXiv:0806.2115

# The Interaction of an 180° Ferroelectric Domain Wall with a Biased Scanning Probe Microscopy Tip: Effective Wall Geometry and Thermodynamics in Ginzburg-Landau-Devonshire Theory

Anna N. Morozovska,[*,1] Sergei V. Kalinin[†,2], Eugene A. Eliseev,[3]
Venkatraman Gopalan,[4] Sergei V. Svechnikov[1]

[1] Institute of Semiconductor Physics, 41, pr. Nauki, 03028 Kiev, and
[3] Institute for Problems of Materials Science, 3, Krjijanovskogo, 03142 Kiev,
National Academy of Science of Ukraine, Ukraine

[2] The Center for Nanophase Materials Sciences and Materials Science and Technology Division, Oak Ridge National Laboratory, Oak Ridge, TN 37831

[4] Department of Materials Science and Engineering,
Pennsylvania State University, University Park, Pennsylvania 16802

## Abstract

The interaction of ferroelectric 180°-domain wall with a strongly inhomogeneous electric field of biased Scanning Probe Microscope tip is analyzed within continuous Landau-Ginzburg-Devonshire theory. Equilibrium shape of the initially flat domain wall boundary bends, attracts or repulses from the probe apex, depending on the sign and value of the applied bias. For large tip-wall separations, the probe-induced domain nucleation is possible. The approximate analytical expressions for the polarization distribution are derived using direct variational method. The expressions provide insight how the equilibrium polarization distribution depends on the wall finite-width, correlation and depolarization effects, electrostatic potential distribution of the probe and ferroelectric material parameters.

[*] Corresponding author: morozo@i.com.ua
[†] Corresponding author: sergei2@ornl.gov

## 1. Introduction

Domain wall motion in disordered media is one of the fundamental mechanisms that control order parameter dynamics in ferroelectric and ferromagnetic materials. The interplay between wall stiffness, driving force, pinning, and thermal excitations gives rise to a broad spectrum of remarkable physical phenomena including transitions between pinned, creep, and sliding regimes, dynamic phase transitions, and self-organized critical behavior. This behavior controlled by *homogeneous* external electric field was studied in details both experimentally and theoretically.[1, 2, 3, 4] In most experimental studies to date, the domain wall dynamics is inferred from the macroscopic response of the system to macroscopic field, detected through changes in polarization, ac susceptibility, lattice parameters, pyroelectric, piezoelectric, and optical properties. Recently, the *local* observations of domain wall geometry and its evolution in uniform external fields have allowed direct information on static wall structure formed after field application, dynamic avalanche time and size distributions, and pinning on individual defects.[5,6,7]

The emergence of the Scanning Probe Microscopy based techniques in the last decade opens the pathway to concentrate electric field within a small (~10-100 nm) volume of material. Combined with electromechanical response detection, this Piezoresponse Force Microscopy approach has been broadly applied for domain imaging and polarization patterning. Piezoresponse force spectroscopy was used to study polarization switching in the small volumes with negligible defect concentration,[8] map distribution of random bond- and random field components of disorder potential,[9] and map polarization switching on a single defect center.[10] These experimental developments have been complemented by the extensive theoretical analysis of domain nucleation mechanisms in the SPM field probe on the ideal surface[11,12,13] and in the presence of charged defects in the rigid ferroelectric approximation (abrupt domain walls).[14] Recently, phase-field and analytical models have emerged to treat this problem in the framework of Landau-Ginzburg-Devonshire (GLD) theory (diffuse walls).[15] Here, we develop the analytical theoretical model for the interaction of the biased SPM probe and 180°-domain wall in the GLD model, paving the way for experimental studies of microscopic mechanisms of domain wall polarization interaction with electric field that can be studied in strongly inhomogeneous fields of biased force microscope probe.

We note that this problem is similar to that of domain wall pinning on a charged impurity, where the SPM probe acts as a "charged impurity" with controlled strength (controlled by tip bias) positioned at a given separation from domain wall. In this context, the problem of the infinitely thin ferroelectric domain wall interaction with a charged point defect was considered by Sidorkin[16]; however neither correlation effects (e.g. finite intrinsic width of domain walls) nor rigorous depolarization field influence were taken into account. For the description of domain wall equilibrium position the Laplace tension conception (whose applicability to ferroelectrics has not been studied in detail) was used instead of the conventional LGD theory, thermodynamic Miller-Weinreich approach[1] or their combination with molecular dynamics and Monte-Carlo simulations as proposed by Rappe et al. [17] However, they studied domain wall profile changes in homogeneous external field.

In this paper, we consider the interaction of ferroelectric 180°-domain wall polarization with a strongly inhomogeneous electric field of biased force microscope probe within LGD thermodynamic approach. The non-linear problem is resolved using direct variational method. The paper is organized as following. The problem statement and basic equations are presented in Section 2. In Section 3 we calculate the influence of the domain wall finite-width, correlation and depolarization effects and ferroelectric material parameters on the equilibrium domain wall profile in the vicinity of biased probe. Coercive field for wall motion and domain nucleation is considered in Section 4. The results and implications for PFM studies of domain walls are discussed in Section 5.

## 2. The problem statement and basic equations

Here we consider the ferroelectric sample region that contains 180°-domain wall positioned in the vicinity of charged force microscope probe [Fig. 1]. The region is free of lattice defects. Maxwell's equations for the inner electric field $\mathbf{E}(\mathbf{r},t) = -\nabla\varphi(\mathbf{r},t)$, expressed via electrostatic potential $\varphi(\mathbf{r},t)$ and polarization $\mathbf{P}(\mathbf{r},t)$ with boundary conditions are:

$$\begin{cases} div(\mathbf{P}(\mathbf{r},t) - \varepsilon_0 \nabla \varphi(\mathbf{r},t)) = 0, & z \geq 0, \\ \varphi(x,y,z=0,t) = V_e(x,y,t), & \varphi(x,y,z=h,t) = 0 \end{cases} \quad (1)$$

Potential distribution $V_e(x,y,t)$ is created by the biased probe of force microscope. The probe is assumed to be in perfect electric contact with the sample surface. Electrostatic potential $\varphi(\mathbf{r},t)$ includes the effects of depolarization field created by polarization bound charges; $\varepsilon_0$ is the dielectric constant, $h$ is the film thickness. The perfect screening of depolarization field outside the sample is realized by the ambient charges, as shown in Fig. 1 (a).

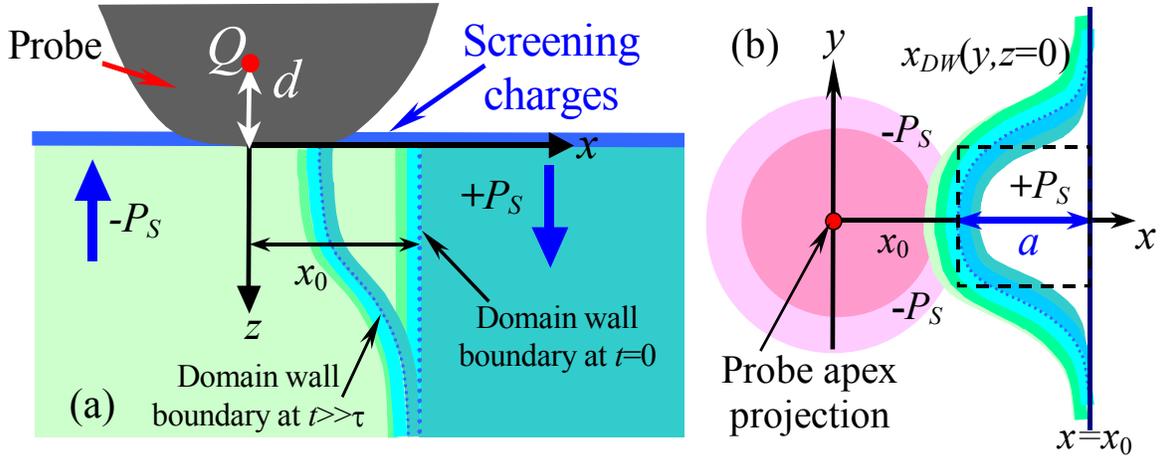

**FIG. 1**. (a) Schematics of ferroelectric 180°-domain wall boundary curved by the strong localized electric field of the biased probe in contact with the sample surface. (b) Wall curvature at the sample surface in quasi-continuous media approximation (solid curves and color-scale). Dashed rectangle corresponds to the schematics of activation field calculations used by Miller and Weinreich[1] for rigid polarization model, where the distance $a$ is equal to the lattice constant.

The polarization $P_3(\mathbf{r},t)$ in uniaxial ferroelectrics is directed along the polar axis, $z$. The sample is dielectrically isotropic in transverse directions, i.e. permittivities $\varepsilon_{11} = \varepsilon_{22}$. The dependence of in-plane polarization components on electric field is linearized as $P_{1,2} \approx -\varepsilon_0(\varepsilon_{11} - 1)\partial \varphi(\mathbf{r})/\partial x_{1,2}$. We can rewrite the problem (1) for quasi-static electrostatic potential as:

$$\begin{cases} \varepsilon_{33}^b \dfrac{\partial^2 \varphi}{\partial z^2} + \varepsilon_{11}\left(\dfrac{\partial^2 \varphi}{\partial x^2} + \dfrac{\partial^2 \varphi}{\partial y^2}\right) = \dfrac{1}{\varepsilon_0}\dfrac{\partial P_3}{\partial z}, \\ \varphi(x,y,z=0) = V_e(x,y,t), \quad \varphi(x,y,z=h) = 0 \end{cases} \qquad (2)$$

Here we introduced dielectric permittivity of background[18] or reference state[19] as $\varepsilon_{33}^b$. Typically $\varepsilon_{33}^b \leq 10$; its origin can be related with electronic polarizability and/or reorientation of impurity dipoles.[20]

The corresponding Fourier-Laplace representation on transverse coordinates $\{x,y\}$ and time $t$ of electric field normal component $\tilde{E}_3(\mathbf{k},z,f) = -\partial\tilde{\varphi}/\partial z$ is the sum of external ($e$) and depolarization ($d$) fields:

$$\tilde{E}_3(\mathbf{k},z,f) = \tilde{E}_3^e(V_e,\mathbf{k},z,f) + \tilde{E}_3^d(P_3,\mathbf{k},z,f), \qquad (3a)$$

$$\tilde{E}_3^e(\mathbf{k},z,f) = \tilde{V}_e(\mathbf{k},f)\frac{\cosh(k(h-z)/\gamma_b)}{\sinh(kh/\gamma_b)}\frac{k}{\gamma_b}, \qquad (3b)$$

$$\tilde{E}_3^d(P_3,\mathbf{k},z,f) = \begin{pmatrix} \displaystyle\int_0^z dz' \dfrac{\tilde{P}_3(\mathbf{k},z',f)}{\varepsilon_0\varepsilon_{33}^b}\cosh(kz'/\gamma_b)\dfrac{\cosh(k(h-z)/\gamma_b)}{\sinh(kh/\gamma_b)}\dfrac{k}{\gamma_b} + \\ \displaystyle\int_z^h dz' \dfrac{\tilde{P}_3(\mathbf{k},z',f)}{\varepsilon_0\varepsilon_{33}^b}\cosh(k(h-z')/\gamma_b)\dfrac{\cosh(kz/\gamma_b)}{\sinh(kh/\gamma_b)}\dfrac{k}{\gamma_b} - \dfrac{\tilde{P}_3(\mathbf{k},z,f)}{\varepsilon_0\varepsilon_{33}^b} \end{pmatrix} \qquad (3c)$$

Here $\gamma_b = \sqrt{\varepsilon_{33}^b/\varepsilon_{11}}$ is the "bare" dielectric anisotropy factor, $\mathbf{k} = \{k_1,k_2\}$ is a spatial wave-vector, its absolute value $k = \sqrt{k_1^2 + k_2^2}$, $f$ is temporal frequency of Laplace transformation. Corresponding Fourier-Laplace image of polarization is $\tilde{P}_3(\mathbf{k},z,f) = \dfrac{1}{2\pi}\int_0^\infty dt \int_{-\infty}^{\infty} dx \int_{-\infty}^{\infty} dy \exp(ik_1x + ik_2y - ft)P_3(x,y,z,t)$. $\tilde{V}_e(\mathbf{k},f)$ is the Fourier-Laplace image of electric field potential at the sample surface. For a transversally homogeneous media, $\varepsilon_{33}^b = 1$ and static case Eq. (3c) reduces to the expression for depolarization field obtained by Kretschmer and Binder.[21]

In the framework of LGD phenomenology, the spatial-temporal evolution of the polarization component $P_3$ of the second order ferroelectric is described by the Landau-Khalatnikov equation:

$$-\tau \frac{d}{dt} P_3 = \alpha P_3 + \beta P_3^3 - \xi \frac{\partial^2 P_3}{\partial z^2} - \eta \left( \frac{\partial^2 P_3}{\partial x^2} + \frac{\partial^2 P_3}{\partial y^2} \right) - E_3 . \qquad (4)$$

The gradient terms $\xi > 0$ and $\eta > 0$, expansion coefficients $\delta > 0$, while $\beta < 0$ for the first order phase transitions or $\beta > 0$ for the second order ones, $\tau$ is the Khalatnikov coefficient (relaxation time). In the absence of (microscopic) pinning centers or for weak pinning of viscous friction type the domain wall equilibrium profile can be found as stationary solution of Eqs. (4). Rigorously, coefficient $\alpha$ should be taken as renormalized by the elastic stress as $(\alpha - 2Q_{ij33}\sigma_{ij})$.[22, 23] Hereinafter we neglect the striction effects, which are relatively small for LiTaO$_3$ and LiNbO$_3$.[24]

Initial and boundary conditions for polarization in Eq. (4) are

$$P_3(\mathbf{r}, t \le 0) = P_0(\mathbf{r}), \qquad \left. \left( P_3 - \lambda_1 \frac{\partial P_3}{\partial z} \right) \right|_{z=0} = 0, \qquad \left. \left( P_3 + \lambda_2 \frac{\partial P_3}{\partial z} \right) \right|_{z=h} = 0. \qquad (5)$$

Here $P_0(\mathbf{r})$ is the initial profile of domain wall that satisfies Eq. (4) for zero external field. The extrapolation lengths $\lambda_{1,2}$ may be different for $z=0$ and $z=h$, reflecting the difference in boundary conditions (e.g. free surface and ferroelectric-electrode interface for thin film, or dissimilar electrodes for capacitor structure or thin dielectric layer on the surface). Reported extrapolation length values are 0.5-50 nm.[25]

Equations (4)-(5) are the closed-form 3D-boundary problem for the determination of the equilibrium domain wall profile. The free energy excess related with the polarization redistribution caused by the external electric field $E_3^e$ can be defined as the energy difference $\Delta G$ between the initial state free energy $G(P_0, E_3^e)$ and the final state free energy $G(P_3, E_3^e)$ with equilibrium polarization distribution $P_3(x, y, z)$ found from Eq.(4):

$$\Delta G(P_3, E_3^e) = G(P_0, E_3^e) - G(P_3, E_3^e), \qquad (6)$$

$$G(P, E_3^e) = \int_{-\infty}^{\infty} dx \int_{-\infty}^{\infty} dy \left( \begin{array}{l} \int_0^h dz \left( \frac{\alpha}{2} P^2 + \frac{\beta}{4} P^4 + \frac{\xi}{2} \left( \frac{\partial P}{\partial z} \right)^2 + \frac{\eta}{2} (\nabla_\perp P)^2 - P \left( E_3^e + \frac{E_3^d}{2} \right) \right) + \\ + \frac{\xi}{2\lambda_1} P^2(z=0) + \frac{\xi}{2\lambda_2} P^2(z=h) \end{array} \right) \qquad (7)$$

In the continuous media approximation, the stable domain wall boundary $x_{DW}(y,z)$ can be defined from the condition $P_3(x_{DW}, y, z, t \gg \tau) = 0$.

For the global excitation of ferroelectric sample with homogeneous external field $E_0^e$, the energetic barrier $\Delta G_a$ required to move the domain wall boundary by overcoming the effect of the lattice discreteness (the simplest pinning model) can be estimated as the difference between the initial state free energy $G(P_0, E_0^e)$ and the equilibrium state $G(P_3, E_0^e)$ with the domain boundary local deviation from initial profile equal to lattice constant $a$.[1] However the simple criteria should be modified for the considered case of probe-induced domain wall bending, since probe-induced domain nucleation far from the wall could appear. The behaviour is analyzed in more detail in Section 4.

For the global excitation of ferroelectric sample with relatively small homogeneous external field, $E_3^e$, the dependence of domain wall velocity **v** upon the electric field usually has exponential form $\mathbf{v} \sim \exp(-E_a/E_0^e)$ in the regime of thermal activation mechanism of domain wall movement.[1] In the Miller-Weinreich model, the critical nucleus size determines the activation energy and thus activation field, $E_a$.

To determine the velocity $\mathbf{v}(x, y, z, t)$ of the domain wall movement far from the activated regime (i.e. in the very large field limit) one can use the substitution $dP_3(\mathbf{r} - \mathbf{v}t)/dt = -\mathbf{v}\nabla P_3$ and corresponding equation for the order parameter.[26] Keeping in mind that the right-hand-side of Eqs.(4) is the free energy (6b) variation derivative $\partial G(P_3)/\partial P_3$, one obtains

$$v_i(x, y, z, t) = \frac{\partial G(P_3)/\partial P_3}{\tau(\partial P_3/\partial x_i)}. \qquad (i=1,2,3) \qquad (8)$$

It is clear that the velocity tends to zero in thermodynamic equilibrium $\partial G(P_3)/\partial P_3 = 0$, as anticipated. Far from the equilibrium variation derivative $\partial G(P_3)/\partial P_3$ can be regarded as generalized pressure similarly to the pressure introduced in the rigid model for domain nucleation as considered by Molotskii et al.[27] Below, we proceed with the analysis of the domain wall geometry and thermodynamics as a key component in the analysis of wall dynamics. The effects of lattice and defect pinning on wall dynamics will be analyzed elsewhere.

## 3. Thermodynamics of domain wall interaction with biased probe

### *3.1. Direct variational method*

Hereinafter, we consider semi-infinite second order ferroelectrics with large extrapolation length, $\lambda_1 \gg \sqrt{\xi}$. Infinite extrapolation length $\lambda_1 \to \infty$ corresponds to the situation of perfect atomic surface structure without defects or damaged layer. Corresponding surface energy proportional to $\frac{\xi}{2\lambda_1}P_3^2(z=0)$ is negligibly small, and hence the domain wall energy is determined only by correlation term, $\frac{\xi}{2}\left(\frac{\partial P_3}{\partial z}\right)^2 + \frac{\eta}{2}(\nabla_\perp P_3)^2$. Details of calculations for much more cumbersome case of finite extrapolation length $\lambda_{1,2} < \infty$ and sample thickness $h < \infty$ are available in Appendices A, B.

Potential distribution produced by the SPM probe on the surface is approximated as $V_e(x,y,t \geq 0) \approx V d/\sqrt{x^2+y^2+d^2}$, where $V$ is the applied bias, $d$ is the effective distance determined by the probe geometry [13, 14]. The potential is normalized assuming the condition of perfect electrical contact with the surface, $V_e(0,0,t \geq 0) \approx V$. The corresponding Fourier-Laplace image for a point-charge approximation of a probe is

$$\tilde{V}_e(\mathbf{k},f) = V\frac{\tilde{w}(\mathbf{k})}{f}, \qquad \tilde{w}(\mathbf{k}) = \frac{d}{k}\exp(-kd). \qquad (9)$$

In the case of local point charge model, the probe is represented by a single charge $Q = 2\pi\varepsilon_0\varepsilon_e R_0 V(\kappa+\varepsilon_e)/\kappa$ located at $d = \varepsilon_e R_0/\kappa$ for a spherical tip apex with curvature $R_0$ ($\kappa \approx \sqrt{(\varepsilon_{33}^b - 1/2\varepsilon_0\alpha)\varepsilon_{11}}$ is the effective dielectric constant determined by the "full" dielectric permittivity in z-direction, $\varepsilon_e$ is ambient dielectric constant), or $d = 2R_0/\pi$ for a flattened tip represented by a disk of radius $R_0$ in contact with the sample surface[13, 14].

Using the perturbation theory, we search for the solution of Eq. (4) in the form

$$P_3(\mathbf{r},t) = P_0(x) + p(\mathbf{r},t). \qquad (10a)$$

Polarization distribution $P_0(x)$ satisfies Eq.(4) at zero external bias, $V_e=0$. Eq.(4) reduces to $\alpha P_0(x) + \beta P_0(x)^3 - \eta \frac{\partial^2 P_0(x)}{\partial x^2} = 0$. The solution for the initial flat domain wall profile positioned at $x=x_0$ is

$$P_0(x) = P_S \tanh\left((x-x_0)/2L_\perp\right). \qquad (10b)$$

where the correlation length is $L_\perp = \sqrt{-\eta/2\alpha}$, and the spontaneous polarization is $P_S^2 = -\alpha/\beta$.

Since the distribution $P_0(x)$ does not cause depolarization field, the operator $E_3^d[P_0(x) + p(\mathbf{r},t)] = E_3^d[p(\mathbf{r},t)] \neq 0$, i.e. depolarization effect is determined by the wall curvature. Hence, the Eq. (4) with substitution (10) acquires the form:

$$\left(\begin{array}{l} \tau\dfrac{dp}{dt} - \left(2\alpha + 3\beta\left(P_S^2 - P_0(x)^2\right)\right)p + 3\beta P_0(x)p^2 + \beta p^3 \\ -\xi\dfrac{\partial^2 p}{\partial z^2} - \eta\left(\dfrac{\partial^2 p}{\partial x^2} + \dfrac{\partial^2 p}{\partial y^2}\right) \end{array}\right) = E_3[V_e, p]. \qquad (11a)$$

Initial and boundary conditions for perturbation $p(x,y,z,t)$ are

$$p(\mathbf{r}, t \leq 0) = 0, \qquad \left.\frac{\partial p}{\partial z}\right|_{z=0} = 0 \qquad (11b)$$

In continuous media approximation adopted here, in the immediate vicinity of domain wall polarization tends to zero and thus Eq. (11a) could be linearized with respect to deviation $p$ from initial profile $P_0$. Using the method of slow varying amplitudes[28] for the linearized Eq.(11a) with $x$-dependent coefficient, we derived the linearized solution of Eq. (11) as:

$$\tilde{p}(\mathbf{k}, z, f) \approx \varepsilon_{33}^b \varepsilon_0 V \frac{\tilde{w}(k)}{f} \frac{(k^2/\gamma_b^2 - s_1^2)(k^2/\gamma_b^2 - s_2^2)}{s_1 s_2 (s_1^2 - s_2^2)} (s_2 \exp(-s_1 z) - s_1 \exp(-s_2 z)), \qquad (12)$$

Eigenvalues $s_{1,2}(k,f)$ are positive roots of biquadratic equation $(s^2 - k^2/\gamma_b^2)(-2\alpha_S + \eta k^2 - \xi s^2) = -s^2/(\varepsilon_0 \varepsilon_{33}^b)$, namely:

$$s_{1,2}^2 = \frac{1 + (k^2(\xi/\gamma_b^2 + \eta) - 2\alpha_S)\varepsilon_0 \varepsilon_{33}^b}{2\varepsilon_0 \varepsilon_{33}^b \xi} \pm \frac{1}{2\varepsilon_0 \varepsilon_{33}^b \xi}\sqrt{\begin{array}{l}\left(1 + (k^2(\xi/\gamma_b^2 + \eta))\varepsilon_0\varepsilon_{33}^b - 2\alpha_S\right)^2 - \\ -4(\varepsilon_{33}^b \varepsilon_0)^2 \xi(k^2 \eta - 2\alpha_S)k^2/\gamma_b^2\end{array}}, \qquad (13)$$

Note, that depolarization field (terms proportional to $\varepsilon_0$) and correlation effects (terms proportional to $\xi$ and $\eta$) determine the spectrum $s_{1,2}(k,f)$. In Appendix A, we have shown that effective coefficient, $\alpha_S$, is renormalized by finite correlation length and lattice relaxation as:

$$\alpha_S(x_0, L_\perp, f) = \alpha\left(1 - \frac{6L_\perp(L_\perp + d)}{\pi\left((L_\perp + d)^2 + x_0^2\right)}\right) - \frac{\tau f}{2} \quad (14)$$

Solution (12) is valid at small biases $V$, for which the amplitude $p$ remained small in comparison with $P_S$.

To obtain the domain wall profile at arbitrary bias we used direct variational method.[29, 30] In this method, **k**-dependent (i.e. coordinate-dependent) part of linearized solution (12) was used as the trial function in the free energy functional $\Delta G(P_3, E_3^e)$ given by Eqs. (6-7), while the amplitude was treated as a variational parameter $P_V$, whose dimensionality is volts. The consequence of this analysis is that $P_V$ is a single parameter determining wall geometry and contributing to free energy. Hence, system thermodynamics is now described by a single scalar quantity, rather than wall coordinates (much like scalar order parameter in GLD).

Direct integration of $\Delta G(P_3, E_3^e)$ along with Eqs. (9)-(12) allows us to determine the amplitude $P_V$ as the solution of nonlinear algebraic equation [see Appendix C for details]. Allowing for the radial symmetry of normalized probe potential $\widetilde{w}(k)$, after elementary algebraic transformations, we obtained the dependence of the *equilibrium solution* and the free energy functional on the applied bias $V$ and other parameters in the form:

$$P_3(\mathbf{r}) \approx P_0(x) - P_V \int_0^\infty \frac{k^2}{\gamma_b} dk \cdot J_0\left(k\sqrt{x^2 + y^2}\right) \widetilde{w}(k) \frac{(s_2 \exp(-s_1 z) - s_1 \exp(-s_2 z))}{\sqrt{\xi(\eta k^2 - 2\alpha_S)}(s_1^2 - s_2^2)}, \quad (15a)$$

$$\Delta G(P_V) \approx \pi d \sqrt{\frac{\varepsilon_0 \varepsilon_{11}}{-2\alpha_S}} \left(-P_V V + \frac{P_V^2}{2} + \frac{w_3}{3} P_V^3 + \frac{w_4}{4} P_V^4\right), \quad (15b)$$

$$P_V + w_3 P_V^2 + w_4 P_V^3 = V. \quad (15c)$$

Where $J_0$ is Bessel function of zero order, the roots $s_{1,2}$ and coefficient $\alpha_S$ are given by Eqs. (13)-(14) under the condition $f=0$ used hereinafter. Parameters

$$w_3(x_0) = \frac{-3\beta P_S x_0}{\sqrt{(L_\perp + d)^2 + x_0^2}} \frac{\sqrt{-2\alpha_S \varepsilon_{11} \varepsilon_0}}{4\alpha_S^2 (L_\perp + d)}, \qquad w_4 = \frac{\beta \varepsilon_{11} \varepsilon_0}{4\alpha_S^2 (L_\perp + d)^2}. \tag{16}$$

are introduced.

Note, that Eq.(15a) is $\varepsilon_{33}^b$-dependent via $\gamma_b = \sqrt{\varepsilon_{33}^b / \varepsilon_{11}}$ and Eq.(13), while bare dielectric permittivity $\varepsilon_{33}^b$ canceled in Eqs.(15b) and (16), since they contain only the product $\varepsilon_{11} \varepsilon_0$.

Both surface and depth profiles of equilibrium polarization distribution perturbed by the biased probe can be calculated from Eq.(15a), where the bias dependence $P_V(V)$ is given by nonlinear Eq.(15c).

The bias dependence of the amplitude $P_V(V)$ given by cubic Eq. (15c) is shown in Fig. 2 (a, b) for different $x_0$ values and LiNbO$_3$ material parameters. Note, that Eq. (15c) reproduces the main features of the ferroelectric hysteresis far from the domain wall (i.e. bistability between the state with single domain wall and the state with nascent domain is possible at $|x_0| \widetilde{>} d$) as shown in Fig. 2 (b). This is a direct consequence of GLD model (as opposed to rigid ferroelectric model) adopted here.

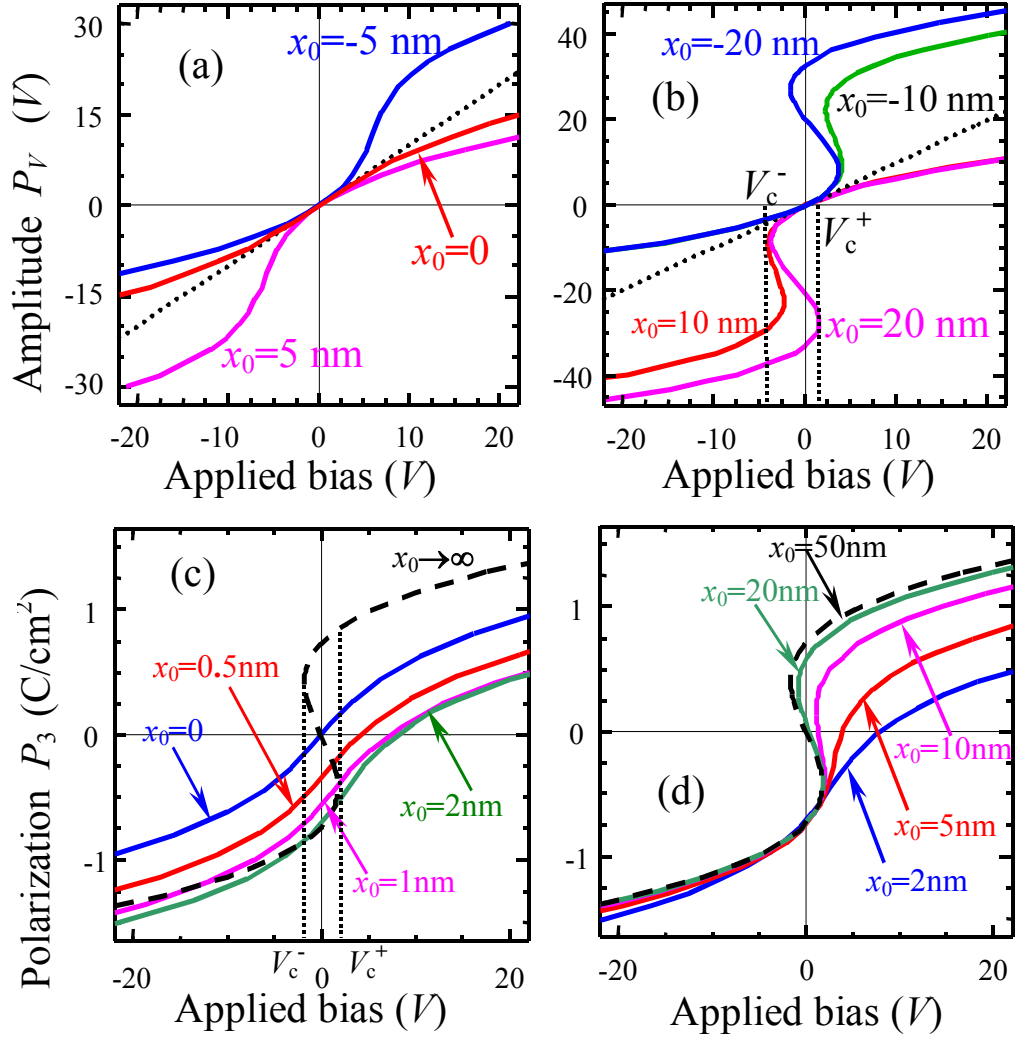

**FIG. 2**. (a, b) Bias dependence of $P_V$ on applied bias $V$ for different domain wall initial position $x_0$ (labels near the curves in nm). Effective distance $d$=5 nm and material parameters for LiNbO$_3$ are $\varepsilon_{11}$=84, $\alpha = -2\cdot10^9$, $\eta=10^{-9}$ in SI units (i.e. $L_\perp$=0.5nm), $P_s$=0.75 C/m$^2$. Dotted curves is linear approximation $P_V=V$ that works satisfactorily up to 5 V for the chosen material parameters. (c, d) polarization below the probe apex, $P_3(0)$, for different $x_0$ values (labels near the curves in nm).

Thermodynamic coercive bias $V_c^\pm$ can then be found from the condition $dV/dP_V = 0$, namely

$$V_c^\pm = \frac{w_3(2w_3^2 - 9w_4) \pm 2(w_3^2 - 3w_4)^{3/2}}{27w_4^2}. \qquad (17)$$

Corresponding hysteresis loop halfwidth, $\Delta V_c = (V_c^+ - V_c^-)/2$, and imprint bias, $V_I = (V_c^+ + V_c^-)/2$, are

$$\Delta V_c = \frac{2(w_3^2 - 3w_4)^{3/2}}{27 w_4^2}, \qquad (18a)$$

$$V_I = \frac{w_3(2w_3^2 - 9w_4)}{27 w_4^2}. \qquad (18b)$$

It is easy to show that $\Delta V_c$ is defined only for the case of $x_0^2 \geq 2(L_\perp + d)^2$. Only in this region the bistability is possible. The coercive biases properties will be considered in details in Section 4. In the next section we show that the hysteresis corresponds to the stable domain formation below the tip apex. At zero bias, the bistable non-zero solutions $P_V(V=0) = \left(-w_3 \pm \sqrt{w_3^2 - 4w_4}\right)/2w_4$ appear under the condition $w_3^2 \geq 4w_4$. The latter inequality is equivalent to condition $x_0^2 \geq 8(L_\perp + d)^2$ [see Appendix C for details].

### 3.2. Equilibrium surface profile of domain wall perturbed by the biased probe

At the sample surface, $z=0$, stationary solution, given by Eq. (15a) at $f=0$, can be simplified as:

$$P_3(x,y,0) = P_0(x) + \int_0^\infty \frac{dk \cdot k^2 \sqrt{\varepsilon_{11}\varepsilon_0} J_0\left(k\sqrt{x^2+y^2}\right)(\eta k^2 - 2\alpha_S)^{-1/2} \tilde{w}(k) \cdot P_V(V)}{\sqrt{1 + \varepsilon_0\left(k^2(\varepsilon_{11}\xi + \eta\varepsilon_{33}^b) - 2\varepsilon_{33}^b \alpha_S\right) + 2\varepsilon_0 k \sqrt{\xi \varepsilon_{11} \varepsilon_{33}^b (\eta k^2 - 2\alpha_S)}}}. \qquad (19)$$

For a point-charge approximation of a probe, $\tilde{w}(k) = d\exp(-kd)/k$ in accordance with Eq. (9). For typical ferroelectric material parameters and $\varepsilon_{33}^b \leq 10$ the inequality $2\varepsilon_0 \varepsilon_{33}^b |\alpha| \ll 1$ is valid, and so the integral in Eq. (19) reduces to the approximate explicit form [see Appendix D for details]:

$$P_3(x,y,0) \approx P_0(x) + \frac{\sqrt{\varepsilon_{11}\varepsilon_0/(-2\alpha_S)}\, d^2 \cdot P_V(V)}{\left(d\sqrt{\eta/(-2\alpha_S)} + d^2 + x^2 + y^2\right)\sqrt{d^2 + x^2 + y^2}}. \qquad (20a)$$

In particular, polarization below the probe apex has the form

$$P_3(\mathbf{r}=0) \approx -P_S \tanh\left(\frac{x_0}{2L_\perp}\right) + \sqrt{\frac{\varepsilon_{11}\varepsilon_0}{-2\alpha_S}} \frac{P_V(V)}{\left(\sqrt{\eta/(-2\alpha_S)} + d\right)}. \qquad (20b)$$

The bias dependence of $P_3(\mathbf{r}=0)$ is illustrated in Figs. 2 (c,d) for different $x_0$ values.

Under the absence of pinning centers, thermodynamically equilibrium domain wall boundary $x_{DW}(y)$ can be determined from the condition $P_3(x_{DW}, y, 0) = 0$. Using expression (10b) for $P_0(x_{DW})$ in Eq. (20), we obtained the parametric dependences

$$x_{DW}(\rho) = x_0 + 2L_\perp \operatorname{arctanh}\left( \frac{-\sqrt{\varepsilon_{11}\varepsilon_0/(-2\alpha_S)} d^2 \cdot P_V/P_S}{\left(d^2 + \rho^2 + \sqrt{\eta/(-2\alpha_S)} d\right)\sqrt{d^2 + \rho^2}} \right), \quad (21a)$$

$$y_{DW}^2(\rho) = \rho^2 - x_{DW}^2(\rho), \quad (21b)$$

valid near the wall ($|x_0| \ll d$). Parameter $\rho$ is the radial coordinate. Far from the wall ($|x_0| > L_\perp + d$) the equilibrium domain appears at biases larger the coercive. The corresponding radius, $\rho(V)$, can be determined from the cubic equation:

$$\left(d^2 + \rho^2 + d\sqrt{\frac{\eta}{-2\alpha_S}}\right)\sqrt{d^2 + \rho^2} = d^2 \sqrt{\frac{\varepsilon_{11}\varepsilon_0}{-2\alpha_S}} \cdot \frac{P_V(V)}{P_S \tanh(x_0/2L_\perp)}. \quad (22)$$

Equilibrium surface profile of domain wall affected by biased probe is shown in Fig. 3 for LiNbO$_3$ material parameters. For chosen polarization distribution the wall attraction to the probe corresponds to positive biases [see Figs.3 (a,c)], while the domain wall repulsion from the probe takes place at negative biases [see Figs.3 (b,d)]. For chosen material constants and probe parameter $d$=5 nm characteristic depth of domain wall bending is close to $d$, as anticipated from Eq.(15a). Note that domain wall boundary bending by biased probe is observed at distances $|x_0| < d$ [see Fig. 4 for $x_0$ = 0, 2, 3 nm], while the probe-induced domain formation appears at $|x_0| > d$ [see Fig. 4 for $x_0$ =5, 7.5 nm].

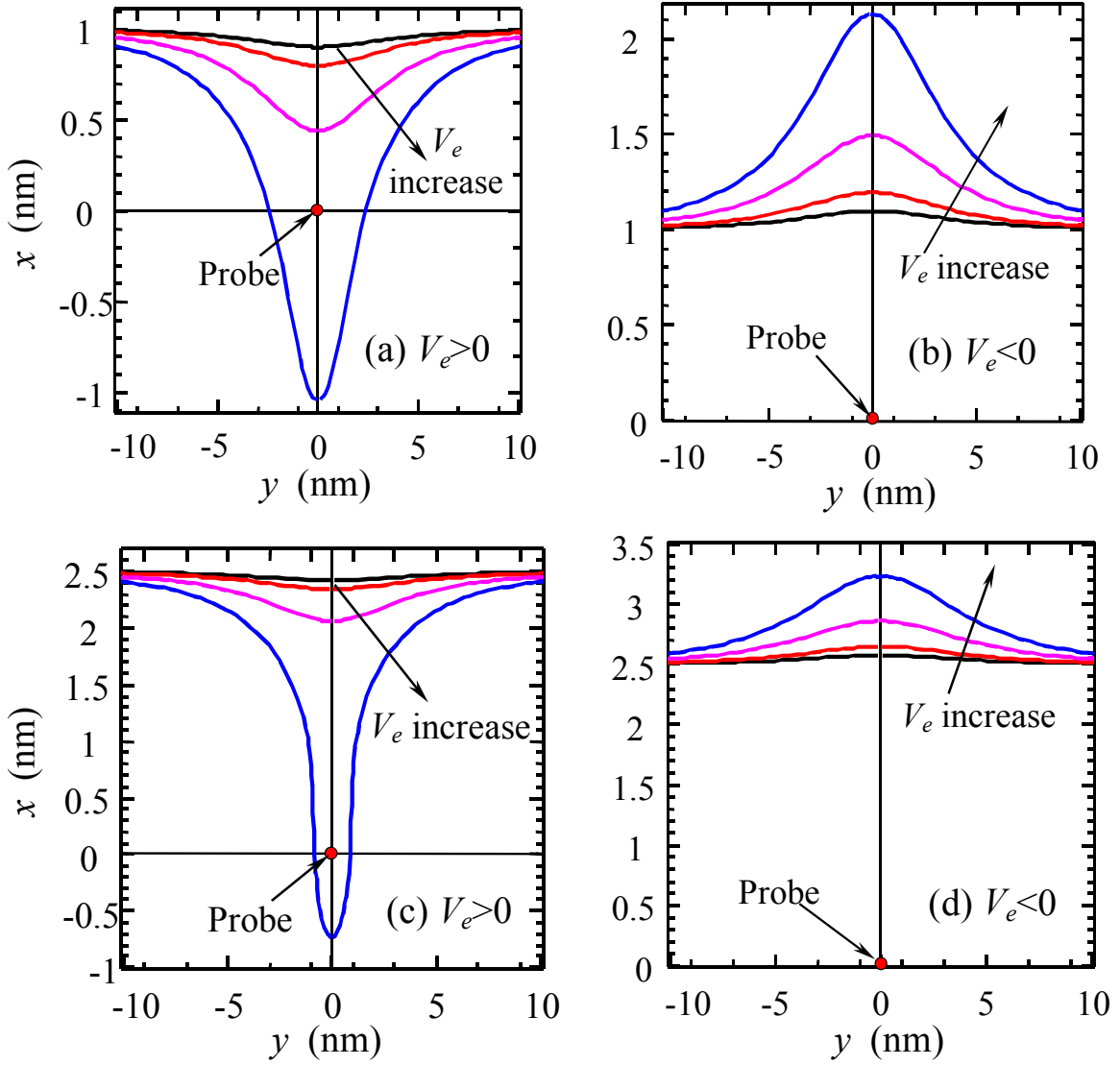

**FIG. 3**. Equilibrium surface profile of domain wall affected by biased probe. Material parameters for LiNbO$_3$ are $\varepsilon_{11}$=84, $\alpha$=-2·10$^9$, $\eta$=10$^{-9}$ in SI units (i.e. $L_\perp$=0.5nm), $P_S$=0.75 C/m$^2$. Effective distance $d$=5 nm and initial domain wall position $x_0$=1 nm for plots (a, b); while $x_0$= 2.5 nm for plots (c, d). Different curves correspond to different voltages applied to the probe: positive $V$ = 1, 2, 5, 10 V (a, c); negative $V$ = -1, -2, -5, -10 V (b, d).

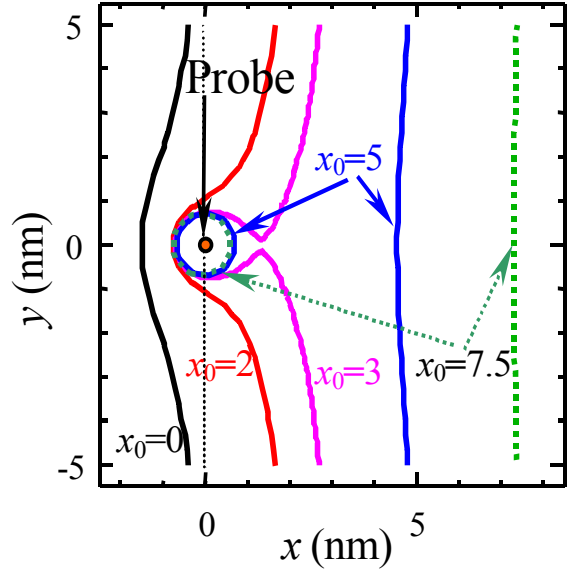

**FIG. 4**. Equilibrium surface profile of domain wall boundary affected by biased probe (for $x_0$=0, 2, 3 nm) and probe-induced domain formation (for $x_0$=5, 7.5 nm). Effective distance $d$=5 nm, applied bias $V$=5V and LiNbO$_3$ material parameters listed in Fig. 3.

**4. Discussion**

Dependence of thermodynamic coercive bias for probe-induced domain formation $V_c^{\pm}$ calculated from Eq. (17) on the distance $x_0$ and $d$ is shown in Figs. 5 (a,b) by dashed and solid curves correspondingly. The presence of $V_c^{\pm}$ indicates the ferroelectric hysteresis appeared in the region $x_0^2 > 2(L_\perp + d)^2$. The asymmetry of $V_c^{\pm}$ corresponds to the domain nucleation and bending towards or away from the tip (compare dashed and solid curves in Figs. 5a). Hysteresis bias changes near the domain wall is due to the fact that the wall can bend towards or away from the tip and its depolarization electric field facilitates or impedes the tip-induced domain nucleation. Shown in Fig. 5b $V_c^{+}$-curves are monotonic, since the tip and depolarization field add together (see dashed curves); $V_c^{-}$-curves have minimum, since tip and depolarization field are opposite, at that depolarization field vanishes far from the wall (see solid curves).

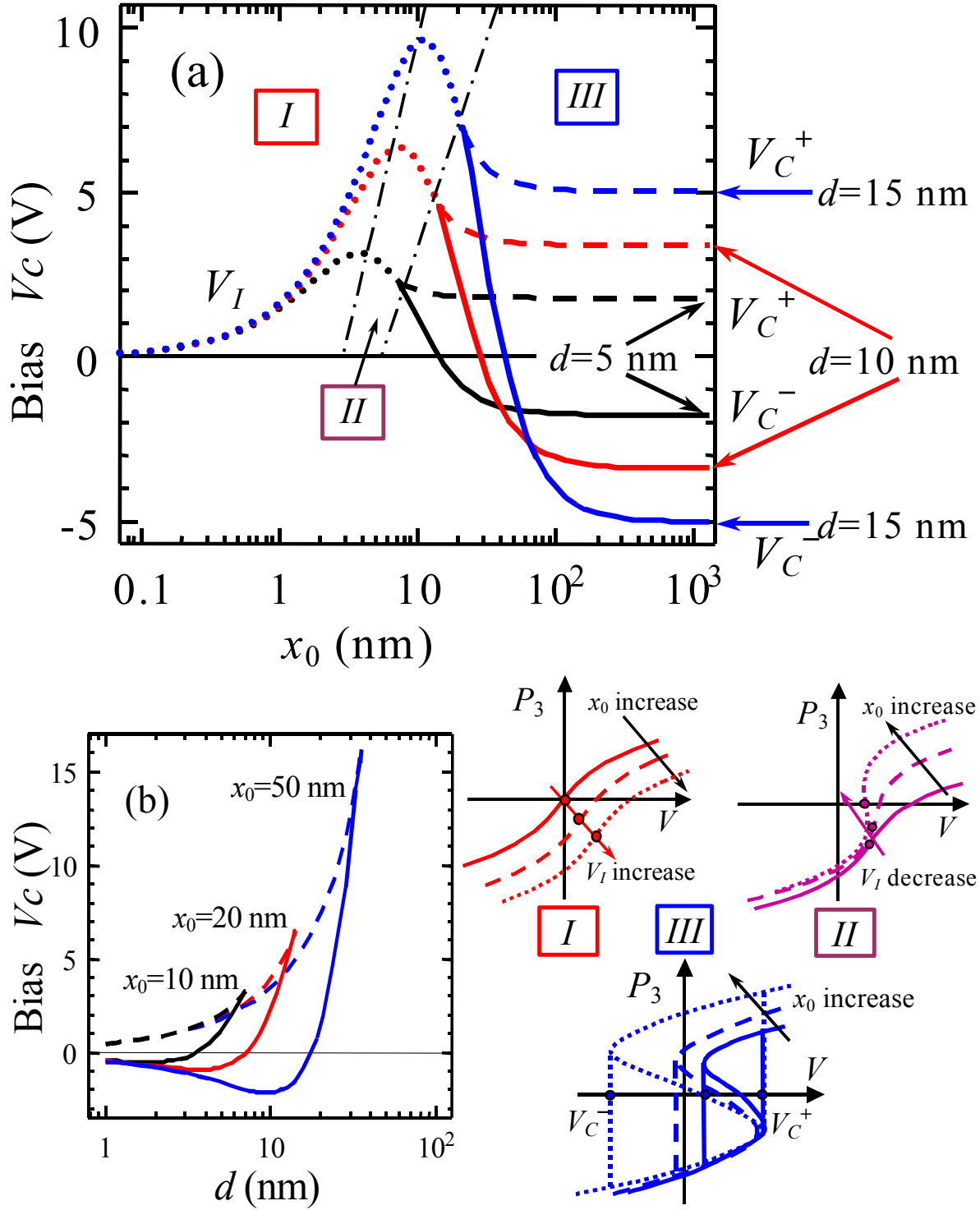

**FIG. 5**. (a) Coercive biases $V_c^{\pm}$ dependence vs. the distance $x_0$ from domain wall for fixed charge-surface separation $d$ and (b) vs. the $d$ values for the fixed distance $x_0$ (labels near the curves). Solid and dashed curves correspond to the left "−" and right "+" values of coercive

field respectively. In the region where hysteresis is absent dotted curves represents the bias $V_I$ at inflection point of polarization dependence on bias. Polarization $P_3$ bias dependence in the regions I-III is schematically shown below at insets I-III. Parameters of LiNbO$_3$ are the same as in Fig.3.

In the region $|x_0| < \sqrt{2}(L_\perp + d)$ hysteresis in absent, only the domain wall bending in different directions depending on the bias sign takes place. The dotted curves correspond to the bias $V_I$ given by Eq.(18b), i.e. the inflection point of polarization dependence on bias, where the second derivative $d^2P_V/dV^2$ is zero. It can be easily shown, that this quantity corresponds to the loop imprint $V_I = (V_c^+ + V_c^-)/2$ in the hysteresis region. Appearance of maximum on $V_I$-curves can be attributed to the linear and nonlinear contributions $w_3$ and $w_4$ of different signs, while both contributions are monotonic functions of $x_0$ [see Eq.(16) and insets I-III]. Physically, the imprint bias originates from nonzero depolarization field (3c) induced by the curved domain wall and nonlinear long-range interactions $\sim P^4$ asymmetry near the wall. Note, that depolarization field is zero for the initial wall profile, $P_0(x)$. From the symmetry considerations, depolarization field and interaction energy are zero when the tip is exactly at the wall, i.e. $x_0=0$. The domain wall bending results in the depolarization field that facilitates domain nucleation in the proximity of the bend, thus reducing local nucleation bias. The interaction energy asymmetry vanishes far from the wall, when the nucleating domain shape becomes axially symmetric. Basically, the peak of the imprint represents, for a tip effective parameter $d$, the intermediate region where the wall influence is seen, but also independent nucleation begins to occur (e.g. as shown in Fig.4 for $x_0=3$ nm).

The dependence of hysteresis loop halfwidth, $\Delta V_c = (V_c^+ - V_c^-)/2$, given by Eq.(18a), and imprint bias $V_I$, given by Eq.(18b), via the domain wall position $x_0$ is shown in Figs. 6 for different $d$ values.

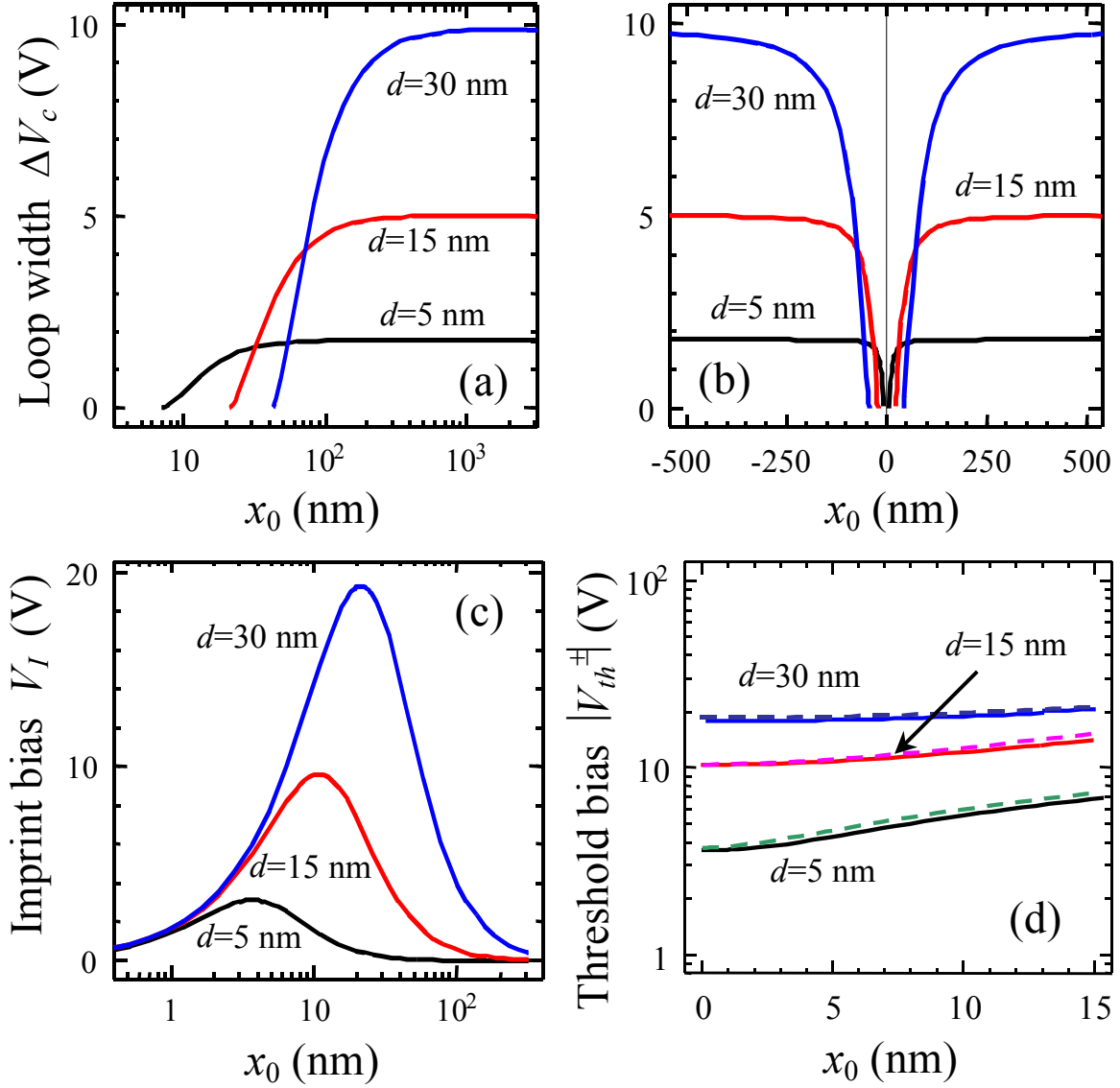

**FIG. 6**. Hysteresis loop width $\Delta V_c$ in linear-log (a) and linear (b) scales; (c) the imprint bias $V_I$ and (d) absolute values of threshold biases $V_{th}^{\pm}$ (solid and dashed curves respectively) as the function of the distance $x_0$ for fixed $d$ values (labels near the curves). Parameters of LiNbO$_3$ are the same as in Fig.3.

As it was mentioned in Section 2, the equilibrium domain wall bending could start at an infinitely small probe bias only in the continuous medium approximation (no lattice or defect pinning). Let us postulate that the threshold (or critical) bias $V_{th}$ is required to move the domain wall boundary by overcoming the effect of the lattice constant discreteness $a$. For the considered case of 1D-initial profile $P_0(x)$, maximal local deviation from the initial profile, $x_{DW} = x_0$, appears at the sample surface [see Fig.1b]. In the activationless region $|x_0| < \sqrt{2}(L_\perp + d)$, absolute values of positive and negative threshold biases, $V_{th}^\pm$, should be found numerically from the condition $\max\{|x_0 - x_{DW}(y = 0, z = 0)|\} = a$ allowing for Eqs.(20a) and (15c). Absolute values of positive and negative threshold bias $V_{th}^\pm$ via $x_0$ are shown in Fig.6d by solid and dashed curves respectively. In the case $|x_0| \ll 2(L_\perp + d)$ amplitude $P_V \approx V$ with high accuracy, allowing for the small value of lattice constant $a$~0.5nm. So, using Eqs. (21a), one leads to the expressions for negative and positive threshold biases [see Appendix D for details]:

$$V_{th}^\pm = \pm \sqrt{\frac{-2\alpha_S}{\varepsilon_{11}\varepsilon_0}} \tanh\left(\frac{a}{2L_\perp}\right) \frac{P_S}{d^2} \left(d^2 + (x_0 \pm a)^2 + L_\perp d\right)\sqrt{d^2 + (x_0 \pm a)^2}. \quad (23)$$

For infinitely thin domain wall $L_\perp \to 0$, and hence Eq.(23) leads to the expression $V_{th}^\pm \to \pm\sqrt{-2\alpha/(\varepsilon_{11}\varepsilon_0)} P_S \left(d^2 + (x_0 \pm a)^2\right)^{3/2}/d^2$.

The phase diagram in coordinates $\{x_0, d\}$ that contains domain wall bending regime (no hysteresis, $V_c^\pm = 0$), to the domain nucleation far from the wall (almost symmetric hysteresis loop with $V_c^+ \approx -V_c^-$) and intermediate regime (asymmetric hysteresis with $V_c^+ \neq -V_c^-$) is shown in Fig. 7.

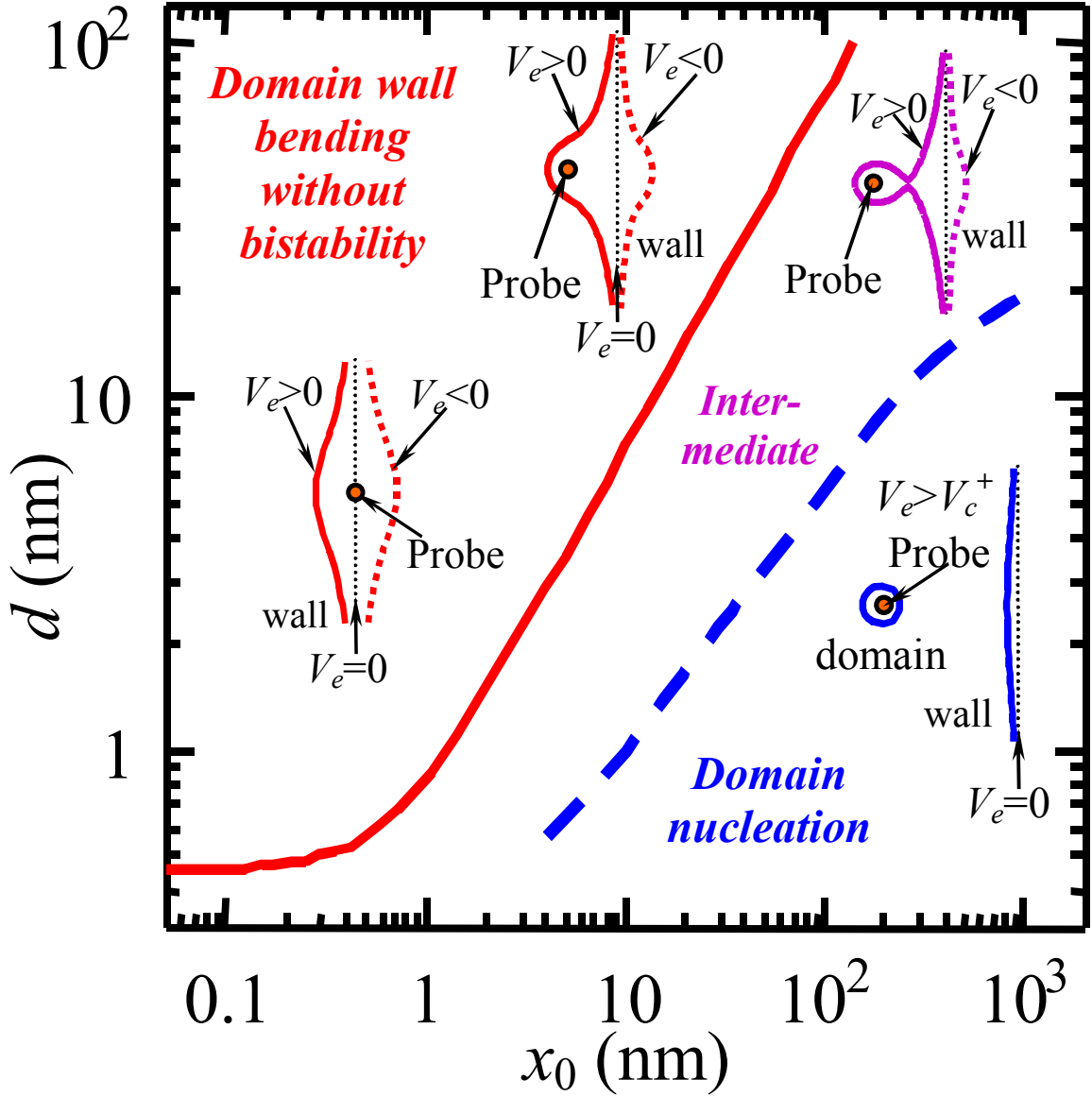

**FIG. 7**. Phase diagram in coordinates $\{x_0, d\}$. Boundaries between the regions of different bias dependence of $P_3$: activationless domain wall bending, domain nucleation far from the wall (hysteresis) and intermediate regimes are shown schematically. Material parameters of LiNbO$_3$ are the same as in Fig.5.

For the second order ferroelectrics considered here, nonzero energetic barrier for polarization reorientation existing in the range of hysteresis, namely at distances $|x_0| > \sqrt{2}(L_\perp + d)$ in the bias range $V_c^- < V < V_c^+$ should be calculated from the free energy (15b) as $E_b(V) = \Delta G(P_V^+, V) - \Delta G(P_V^-, V)$. Free energy $\Delta G(P_V)$ calculated from Eq.(15b) for different bias $V$ are shown in Figs. 8 (a-c). Orientation barriers via the distance $x_0$ from domain wall are shown in Fig. 8 (d) for different charge-surface separation $d$.

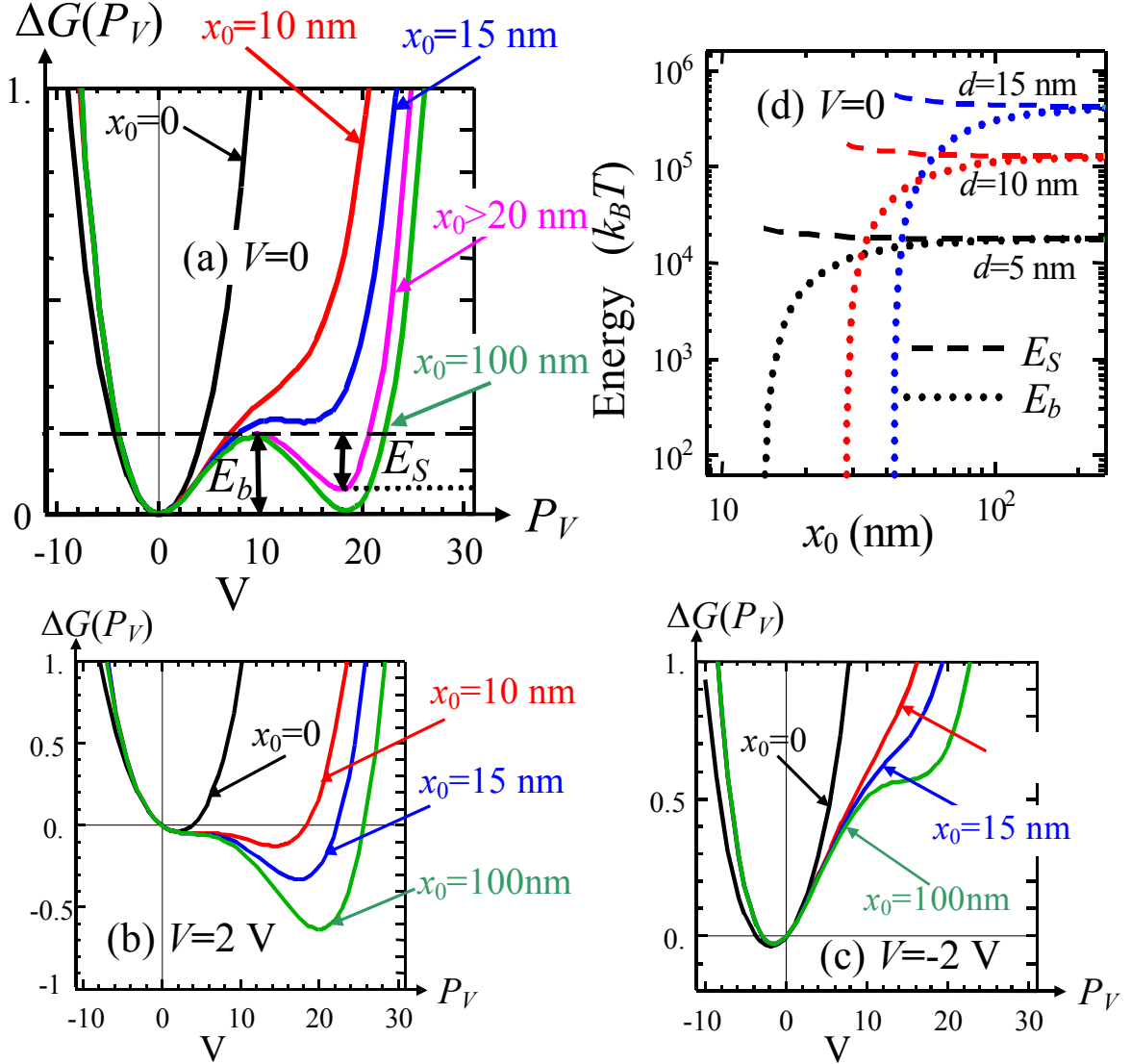

**FIG. 8**. (a, b, c) Free energy $\Delta G(P_V)$ at different bias V=0 (a), 2V (b), -2V(c). (d) Orientation barrier of nascent domain $E_b$ (dashed curves) and metastable saddle point energy $E_S$ (dotted curves) via the distance $x_0$ from domain wall for fixed charge-surface separation $d$ (marked near the curves) and V=0. Material parameters of LiNbO$_3$ are the same as in Fig.5.

Within direct variational method the nonlinear behavior of polarization and was approximately described by substitution of linearized solution of LGD-equation into the free energy with nonlinear terms. This allows us to obtain relatively simple analytical expressions for domain wall surface profile depending on the one variational parameter $P_V$. However, rigorously speaking, one should estimate the accuracy of the one-parametric trial function. Recent comparison of approximate one-parametric trial function based on 2D-linearized solution of LGD-equation with numerical calculations performed by phase-field modeling has shown that the one-parametric trial function works surprisingly well for the description of domain wall surface broadening.[24] This encourages us to use the present one-parametric trial function for the description of the interaction of an 180º-ferroelectric domain wall with a biased tip, its surface bending near the probe and obtained radius of nucleating domain. For more rigorous analytical calculations of polarization depth profile and length of tip-induced domains at least two-parametric trial function may be necessary.

## 5. Summary

We have analyzed in detail the voltage-dependent thermodynamics and geometry of domain wall in the presence of the localized electric field, corresponding to the physical cases of domain wall dynamics on the presence of biased SPM probe. Linearized solution of Landau-Ginzburg-Devonshire equation for wall profile is valid for small tip biases. The direct variational method allowed extending this analytical solution to the strongly non-linear case of arbitrary probe biases, providing the full thermodynamic description of the system in terms of a single voltage-dependent scalar potential. In the uniform field case, corresponding to the infinite tip-surface separation, the potential will become the standard GLD potential function in the uniform field. Obtained analytical expressions provide insight how the equilibrium polarization distribution depends on the wall finite-width, correlation and depolarization effects, electrostatic potential distribution of the probe and ferroelectric material parameters.

Depending on probe parameters and probe-wall separation, the bias dependence of potential can be single valued, corresponding to the activationless domain wall bending. For larger values of the probe-wall separation, the potential can exhibit bistability, corresponding to ferroelectric hysteresis. The switching between polarization direction $+P_3$ and $-P_3$ defines the thermodynamic coercive bias of tip-induced domain switching affected by the domain wall. We demonstrate that for small tip-surface separations the domain wall displacement is activationless, corresponding to the wall bending towards or away from the probe. For intermediate separations, the process is affected by depolarization field induced by wall bending, corresponding to thermodynamics nucleation biases reduced relative to bulk values and appearance of significant loop imprint. Finally, for large tip-surface separations, the wall does not affect nucleation below the tip.

This analysis performed for the case of ferroelectric material with second order phase transition in the absence of lattice and defect pinning. It can further be extended to incorporate lattice effects though the introduction of lattice discreteness or periodic pinning potentials.

**Acknowledgement**

SVK greatly acknowledges multiple discussions with A.K. Tagantsev (EPFL). Research is sponsored in part (SVK) by the Center for Nanophase Materials Sciences, Office of Basic Energy Sciences, U.S. Department of Energy. VG wishes to gratefully acknowledge financial support from the National Science Foundation grant numbers DMR-0602986, 0512165, 0507146, and 0213623, and CNMS at Oak Ridge National Laboratory.

**Appendix A. Linearized equation with x-dependent coefficients**

We can rewrite the problem (1) for quasi-static electrostatic potential as:

$$\begin{cases} \varepsilon_{33}^b \frac{\partial^2 \varphi}{\partial z^2} + \varepsilon_{11}\left(\frac{\partial^2 \varphi}{\partial x^2} + \frac{\partial^2 \varphi}{\partial y^2}\right) = \frac{1}{\varepsilon_0}\frac{\partial P_3}{\partial z} \\ \varphi(x,y,z=0,t) = V_e(x,y,t), \quad \varphi(x,y,z=h) = 0 \end{cases} \quad (A.1)$$

Corresponding Fourier-Laplace representation on transverse coordinates $\{x,y\}$ and time $t$ of electrostatic potential $\tilde{\varphi}(\mathbf{k},z,f)$ and electric field normal component $\tilde{E}_3(\mathbf{k},z,f) = -\partial\tilde{\varphi}/\partial z$ have the form:

$$\tilde{\varphi}(\mathbf{k},z,f) = \tilde{V}_e(\mathbf{k},f)\frac{\sinh(k(h-z)/\gamma_b)}{\sinh(kh/\gamma_b)} + \begin{pmatrix} \int_0^z dz'\, \tilde{P}_3(\mathbf{k},z',f)\frac{\cosh(kz'/\gamma_b)\sinh(k(h-z)/\gamma_b)}{\varepsilon_0 \varepsilon_{33}^b \cdot \sinh(kh/\gamma_b)} - \\ \int_z^h dz'\, \tilde{P}_3(\mathbf{k},z',f)\frac{\sinh(kz/\gamma_b)\cosh(k(h-z')/\gamma_b)}{\varepsilon_0 \varepsilon_{33}^b \cdot \sinh(kh/\gamma_b)} \end{pmatrix}$$

(A.2)

$\gamma_b = \sqrt{\varepsilon_{33}^b/\varepsilon_{11}}$ is the "bare" dielectric anisotropy factor.

Fourier image on transverse coordinates $\{x,y\}$ and Laplace on time $t$ of linearized Eq.(11a) for $\tilde{p}(f,\mathbf{k},z) = \frac{1}{2\pi}\int_0^\infty dt \int_{-\infty}^\infty dx \int_{-\infty}^\infty dy\, \exp(ik_1 x + ik_2 y - ft)p(t,x,y,z)$ gives

$$(\tau f - 2\alpha)\tilde{p}(f,\mathbf{k},z) - \left(\xi\frac{d^2}{dz^2} - \eta k^2\right)\tilde{p}(f,\mathbf{k},z) = \begin{pmatrix} \tilde{E}_3(p,f,\mathbf{k},z) + \\ + 3\beta P_S^2 \int_{-\infty}^\infty \frac{d\tilde{k}_1}{2\pi} F(\tilde{k}_1)\tilde{p}(f,k_1-\tilde{k}_1,k_2,z) \end{pmatrix} \quad (A.3)$$

Where the function $F(k_1) = \frac{4\pi k_1 L_\perp^2 \exp(ik_1 x_0)}{\sinh(\pi k_1 L_\perp)}$ is Fourier-Laplace image of $P_S^2 - P_0(x)^2 = P_S^2/\cosh^2((x-x_0)/2L_\perp)$, integrated over $k_2$ and normalized on $P_S^2$.

Let us apply operator $\varepsilon_{33}^b \frac{d^2}{dz^2} - \varepsilon_{11} k^2$ to the electric field $\tilde{E}_3(\mathbf{k},z)$. After simple but cumbersome transformations one obtains that $\left(\varepsilon_{33}^b \frac{d^2}{dz^2} - \varepsilon_{11}k^2\right)\tilde{E}_3(\mathbf{k},z) = -\frac{1}{\varepsilon_0}\frac{d^2 \tilde{P}_3(\mathbf{k},z)}{dz^2}$ as anticipated directly from Eq.(A.1). Then applying operator $\varepsilon_{33}^b \dfrac{d^2}{dz^2} - \varepsilon_{11} k^2$ to linearized Eq. (A.3), we obtained

$$\left(\left(\varepsilon_{33}^b \dfrac{d^2}{dz^2} - \varepsilon_{11} k^2\right)\left((\tau f - 2\alpha) - \left(\xi \dfrac{d^2}{dz^2} - \eta k^2\right)\right) + \dfrac{1}{\varepsilon_0} \dfrac{d^2}{dz^2}\right) \tilde{p}(f,\mathbf{k},z) =$$
$$= 3\beta P_S^2 \left(\varepsilon_{33}^b \dfrac{d^2}{dz^2} - \varepsilon_{11} k^2\right) \int_{-\infty}^{\infty} \dfrac{d\tilde{k}_1}{2\pi} F(\tilde{k}_1) \cdot \tilde{p}(f, k_1 - \tilde{k}_1, k_2, z)$$
(A.4)

Since function $F$ has maximum at $\tilde{k}_1 = 0$ and the perturbation $\tilde{p}$ should vanish in **k**-space as the probe electric field, i.e. $\tilde{p} \sim \exp\left(-d\sqrt{k_1^2 + k_2^2}\right)$, one may estimate

$$\int_{-\infty}^{\infty} \dfrac{d\tilde{k}_1}{2\pi} F(\tilde{k}_1) \tilde{p}(f, k_1 - \tilde{k}_1, k_2, z) \approx c(x_0, d, L_\perp) \tilde{p}(f,\mathbf{k},z)$$
$$c(x_0, d, L_\perp) = \dfrac{1}{\pi^2}\left(\text{PolyGamma}\left(1, \dfrac{d + \pi L_\perp - ix_0}{2\pi L_\perp}\right) + \text{PolyGamma}\left(1, \dfrac{d + \pi L_\perp + ix_0}{2\pi L_\perp}\right)\right)$$
(A.5)

Here PolyGamma($n$, $z$) gives the $n^{th}$ derivative of the digamma function $\psi^{(n)}(z) = d^n\psi(z)/dz^n$. PolyGamma($z$) is the logarithmic derivative of the gamma function, given by $\psi(z) = \Gamma'(z)/\Gamma(z)$.

With a good degree of approximation, $F(k_1) \sim 2L_\perp \exp(ik_1 x_0 - \pi |k_1| L_\perp)$, so the convolution in Eq.(A.5) can be estimated as Pade approximation $\dfrac{4L_\perp (L_\perp + d)}{\pi\left((L_\perp + d)^2 + x_0^2\right)} \tilde{p}(f, \mathbf{k}, z)$ [see Fig. 9].

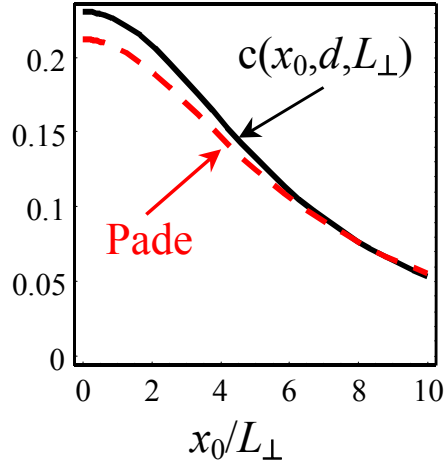

**FIG. 9**. Dimensionless function $c(x_0, d, L_\perp)$ appeared in Eq.(A.5) (solid curve) and its Pade approximation $\dfrac{4L_\perp (L_\perp + d)}{\pi\left((L_\perp + d)^2 + x_0^2\right)}$ (dashed curve) for parameter $d/L_\perp = 5$.

Finally, Eq.(A.2) acquires the form:

$$\left(\left(\varepsilon_{33}^b \frac{d^2}{dz^2} - \varepsilon_{11} k^2\right)\left(\tau f - 2\alpha\left(1 - \frac{6L_\perp (L_\perp + d)}{\pi\left((L_\perp + d)^2 + x_0^2\right)}\right) - \xi\frac{d^2}{dz^2} + \eta k^2\right) + \frac{1}{\varepsilon_0}\frac{d^2}{dz^2}\right)\tilde{p}(f,\mathbf{k},z) = 0$$

(A.6)

Thus "effective" coefficient could be introduced as

$\alpha_S(x_0, L_\perp, f) = \alpha\left(1 - \dfrac{6L_\perp(L_\perp + d)}{\pi\left((L_\perp + d)^2 + x_0^2\right)}\right) - \dfrac{\tau f}{2}$. The boundary conditions

$$\left(\tilde{p}(f,\mathbf{k},z) - \lambda_1 \frac{d\tilde{p}(f,\mathbf{k},z)}{dz}\right)\bigg|_{z=0} = 0, \quad \left(\tilde{p}(f,\mathbf{k},z) + \lambda_2 \frac{d\tilde{p}(f,\mathbf{k},z)}{dz}\right)\bigg|_{z=h} = 0. \quad (A.7)$$

**Appendix B. Solution of linearized equation with "effective" constant coefficients**

Looking for the solution of Eq. (A.6) in the form $\tilde{p}(f,\mathbf{k},z) \sim \exp(sz)$, one can find characteristic equation for eigenvalues $s(f,k)$ in the form of biquadratic equation:

$$\left(\varepsilon_{33}^b s^2 - \varepsilon_{11} k^2\right)\left(-2\alpha_S - \left(\xi s^2 - \eta k^2\right)\right) = -\frac{s^2}{\varepsilon_0}. \quad (B.1)$$

The roots of Eq.(B.1) are

$$s_{1,2}^2(f,k) = \begin{pmatrix} \dfrac{1+\left(k^2\left(\xi\varepsilon_{11}/\varepsilon_{33}^b+\eta\right)\right)-2\alpha_S\right)\varepsilon_0\varepsilon_{33}^b}{2\varepsilon_0\varepsilon_{33}^b\xi} \\ \pm\dfrac{1}{2\varepsilon_0\varepsilon_{33}^b\xi}\sqrt{\left(1+\left(k^2\left(\xi\varepsilon_{11}/\varepsilon_{33}^b+\eta\right)\right)\varepsilon_0-2\alpha_S\right)^2-4k^2\varepsilon_{11}\varepsilon_{33}^b\varepsilon_0^2\xi\left(k^2\eta-2\alpha_S\right)} \end{pmatrix}$$ (B.2)

It is seen that for any real values of $k$ values of $s_{1,2}(f,k)$ are real and identities are valid:

$$s_1^2 + s_2^2 = \frac{1+\left(-2\alpha_S+k^2\left(\xi/\gamma_b^2+\eta\right)\right)\varepsilon_0\varepsilon_{33}^b}{\varepsilon_0\varepsilon_{33}^b\xi}, \quad s_1^2 s_2^2 = \frac{k^2}{\gamma_b^2}\left(\frac{-2\alpha_S+\eta k^2}{\xi}\right),$$ (B.3a)

$$\frac{k^4}{\gamma_b^4} - \frac{k^2}{\gamma_b^2}\left(s_1^2+s_2^2\right)+s_1^2 s_2^2 = -\frac{k^2}{\varepsilon_0\varepsilon_{33}^b\gamma_b^2\xi}.$$ (B.3b)

In the limiting case $k \to 0$:

$$s_1 \approx \sqrt{\frac{1+\left(-2\alpha_S\right)\varepsilon_0\varepsilon_{33}^b}{\varepsilon_0\varepsilon_{33}^b\xi}}, \qquad s_2 \approx k\sqrt{\frac{\varepsilon_{11}\varepsilon_0\left(-2\alpha_S\right)}{1+\left(-2\alpha_S\right)\varepsilon_0\varepsilon_{33}^b}}.$$ (B.3c)

So that the general solution of Eq. (A.6) acquires the form

$$\tilde{p}(f,\mathbf{k},z) = A_1 \cosh(s_1 z) + B_1 \cosh(s_1(h-z)) + A_2 \cosh(s_2 z) + B_2 \cosh(s_2(h-z))$$ (B.4)

After substitution of Eq.(B.4) into (A.6) one obtains:

$$0 = \frac{\tilde{V}_e(\mathbf{k},f)}{-2\alpha_S}\frac{k\cosh(k(h-z)/\gamma_b)}{\gamma_b \sinh(k h/\gamma_b)} -$$

$$-\frac{1}{\varepsilon_0\varepsilon_{33}^b\left(-2\alpha_S\right)}\frac{k s_1\gamma_b}{k^2-s_1^2\gamma_b^2}\frac{\sinh(s_1 h)}{\sinh(k h/\gamma_b)}\left(A_1\cosh\left(\frac{k}{\gamma_b}z\right)+B_1\cosh\left(\frac{k}{\gamma_b}(h-z)\right)\right)$$ (B.5)

$$-\frac{1}{\varepsilon_0\varepsilon_{33}^b\left(-2\alpha_S\right)}\frac{k s_2\gamma_b}{k^2-s_2^2\gamma_b^2}\frac{\sinh(s_2 h)}{\sinh(k h/\gamma_b)}\left(A_2\cosh\left(\frac{k}{\gamma_b}z\right)+B_2\cosh\left(\frac{k}{\gamma_b}(h-z)\right)\right)$$

Eq.(B.5) along with the boundary conditions (A.7) leads to the system of equations for constants $A_i$ and $B_i$:

$$\frac{k s_1\gamma_b}{k^2-s_1^2\gamma_b^2}\sinh(s_1 h)B_1 + \frac{k s_2\gamma_b}{k^2-s_2^2\gamma_b^2}\sinh(s_2 h)B_2 = \varepsilon_{33}^b\varepsilon_0\tilde{V}_e(\mathbf{k},f)\frac{k}{\gamma_b},$$ (B.6a)

$$\frac{k s_1\gamma_b}{k^2-s_1^2\gamma_b^2}\sinh(s_1 h)A_1 + \frac{k s_2\gamma_b}{k^2-s_2^2\gamma_b^2}\sinh(s_2 h)A_2 = 0,$$ (B.6b)

$$\begin{pmatrix} A_1+B_1\cosh(s_1 h)+A_2+B_2\cosh(s_2 h)+ \\ +\lambda_1 B_1 s_1 \sinh(s_1 h)+\lambda_1 B_2 s_2 \sinh(s_2 h) \end{pmatrix} = -\frac{P_0(\mathbf{k})}{f},$$ (B.6c)

$$\begin{pmatrix} A_1 \cosh(s_1 h) + B_1 + A_2 \cosh(s_2 h) + B_2 + \\ + \lambda_2 A_1 s_1 \sinh(s_1 h) + \lambda_2 A_2 s_2 \sinh(s_2 h) \end{pmatrix} = -\frac{P_0(\mathbf{k})}{f} \quad \text{(B.6d)}$$

Further let us consider the case and $\lambda_1 \to \infty$, $h \to \infty$, when

$$\tilde{p}(f,\mathbf{k},z) = \varepsilon_{33}^b \varepsilon_0 \tilde{V}_e(f,\mathbf{k}) \left( \frac{s_2 \cdot \exp(-s_1 z)}{s_2 M(s_1) - s_1 M(s_2)} + \frac{s_1 \cdot \exp(-s_2 z)}{s_1 M(s_2) - s_2 M(s_1)} \right) \quad \text{(B.7)}$$

Where $M(s) = \dfrac{s}{\varepsilon_{11} k^2 - s^2}$. In the explicit form:

$$\tilde{p}(f,\mathbf{k},z) = \varepsilon_{33}^b \varepsilon_0 \tilde{V}_e(f,\mathbf{k}) \frac{(k^2/\gamma_b^2 - s_1^2)(k^2/\gamma_b^2 - s_2^2)}{s_1 s_2 (s_1^2 - s_2^2)} (s_2 \exp(-s_1 z) - s_1 \exp(-s_2 z)), \quad \text{(B.8a)}$$

$$\tilde{p}(f,\mathbf{k},0) = -\varepsilon_{33}^b \varepsilon_0 \tilde{V}_e(f,\mathbf{k}) \frac{(k^2/\gamma_b^2 - s_1^2)(k^2/\gamma_b^2 - s_2^2)}{s_1 s_2 (s_1 + s_2)}$$

$$\approx \frac{\tilde{V}_e(f,\mathbf{k})}{-2\alpha_S} k \sqrt{\frac{\varepsilon_{11} \varepsilon_0 (-2\alpha_S)}{(1 + (-2\alpha_S)\varepsilon_0 \varepsilon_{33}^b)}} \quad \text{(B.8b)}$$

**Appendix C. Direct variational method for equilibrium polarization distribution**

Equilibrium polarization $P_3(\mathbf{r}) = P_0(x) + p(\mathbf{r})$ should be substituted into the free energy in $r$-space:

$$G(P_3) = \frac{1}{S} \int_{-\infty}^{\infty} dx \int_{-\infty}^{\infty} dy \int_0^h dz \left( \frac{\alpha}{2} P_3^2 + \frac{\beta}{4} P_3^4 + \frac{\xi}{2} \left( \frac{\partial P_3}{\partial z} \right)^2 + \frac{\eta}{2} (\nabla_\perp P_3)^2 - P_3 \left( E_3^e + \frac{E_3^d}{2} \right) \right) \quad \text{(C.1a)}$$

$S$ is the sample cross-section. In $\mathbf{k}$-representation for $\tilde{P}_3(\mathbf{k},z)$:

$$G(\tilde{P}_3) = \frac{1}{S} \int_0^h dz \int_{-\infty}^{\infty} dk_1 \int_{-\infty}^{\infty} dk_2 \left( \frac{\alpha}{2} |\tilde{P}_3|^2 + \frac{\beta}{4} |\tilde{P}_3^2|^2 + \frac{\xi}{2} \left| \frac{\partial \tilde{P}_3}{\partial z} \right|^2 + \frac{\eta}{2} k^2 |\tilde{P}_3|^2 - \tilde{P}_3 \left( \tilde{E}_3^{*e} + \frac{\tilde{E}_3^{*d}}{2} \right) \right) \quad \text{(C.1b)}$$

Where external field $E_3^e(x,y,z) \sim V$. We neglect the surface energy, assuming $\lambda_{1,2} \to \infty$. In $r$-space the trial function parts $P_0(x)$ satisfy nonlinear equation

$$\alpha P_0(x) + \beta P_0(x)^3 - \eta \frac{\partial^2 P_0(x)}{\partial x^2} = 0 \quad \text{(C.2a)}$$

Perturbation $p(x,y,z) = P_V \cdot q(x,y,z)$ satisfy linearized equation with boundary conditions

$$\left(\alpha + 3\beta P_0(x)^2\right)p - \xi\frac{\partial^2 p}{\partial z^2} - \eta\left(\frac{\partial^2 p}{\partial x^2} + \frac{\partial^2 p}{\partial y^2}\right) = E_3^V + E_3^d[p], \tag{C.3a}$$

$$\frac{\partial p(z=0)}{\partial z} = 0, \qquad \frac{\partial p(z=h)}{\partial z} = 0. \tag{C.3b}$$

Here $E_3^V(x,y,z) = \frac{P_V}{V}E_3^e(x,y,z)$. Allowing for the boundary conditions (C.3b), electrical boundary conditions $\int_0^h dz\, E_3 = V_e(x,y)$, nonlinear equation of state (11a) for exact solution could be rewritten as $\int_0^h dz\left(\left(\alpha + 3\beta P_0(x)^2\right)p - \eta\left(\frac{\partial^2 p}{\partial x^2} + \frac{\partial^2 p}{\partial y^2}\right) + 3\beta P_0(x)p^2 + \beta p^3\right) = V_e(x,y)$.

Let us find the variational amplitude $P_V$ of the trial function $p(x,y,z) = P_V \cdot q(x,y,z)$ to satisfy the boundary conditions (C.3b) and the nonlinear equation at least in average, i.e. from the minimum of the free energy (C.1). Taking into account that $\int_0^h dz\, E_3^d = 0$, we obtained from (C.1) that

$$\Delta G = \frac{1}{S}\int_{-\infty}^{\infty} dx \int_{-\infty}^{\infty} dy \int_0^h dz \begin{pmatrix} \frac{\alpha}{2}(p^2 + 2P_0 p) + \frac{\beta}{4}(p^4 + 4P_0 p^3 + 6P_0^2 p^2 + 4P_0^3 p) \\ + \frac{\xi}{2}\left(\frac{\partial p}{\partial z}\right)^2 + \frac{\eta}{2}\left(\left(\frac{\partial p}{\partial x}\right)^2 + \left(\frac{\partial p}{\partial y}\right)^2 + 2\left(\frac{\partial p}{\partial x}\right)\left(\frac{\partial P_0}{\partial x}\right)\right) \\ - p\left(E_3^e + \frac{1}{2}E_3^d[p]\right) \end{pmatrix}. \tag{C.4}$$

Substituting depolarization field as $E_3^d[p] = \left(\alpha + 3\beta P_0^2(x)\right)p - \xi\frac{\partial^2 p}{\partial z^2} - \eta\left(\frac{\partial^2 p}{\partial x^2} + \frac{\partial^2 p}{\partial y^2}\right) - E_3^e\frac{P_V}{V}$, elementary transformations and integration over parts $\int_0^h dz\left(\left(\frac{\partial p}{\partial z}\right)^2 + p\frac{\partial^2 p}{\partial z^2}\right) = 0$,

$\int_{-\infty}^{\infty} dx \int_{-\infty}^{\infty} dy\left(\left(\frac{\partial p}{\partial x}\right)^2 + \left(\frac{\partial p}{\partial y}\right)^2 + p\left(\frac{\partial^2 p}{\partial x^2} + \frac{\partial^2 p}{\partial y^2}\right)\right) = 0$ leads to the free energy expression

$$\Delta G(P_V) = \frac{1}{S}\int_{-\infty}^{\infty} dx \int_{-\infty}^{\infty} dy \int_0^h dz\left((\alpha P_0 + \beta P_0^3 - E_3^e)p + \eta\left(\frac{\partial p}{\partial x}\right)\left(\frac{\partial P_0}{\partial x}\right) + \frac{E_3^e P_V}{2V}p + \beta P_0 p^3 + \frac{\beta}{4}p^4\right).$$

Allowing for Eq.(C.2a) and integration over parts $\left(\frac{\partial p}{\partial x}\right)\left(\frac{\partial P_0}{\partial x}\right) \to -p\frac{\partial^2 P_0}{\partial x^2}$, we obtained for $p = P_V \cdot q$:

$$\Delta G(P_V) = \frac{1}{S}\int_{-\infty}^{\infty} dx \int_{-\infty}^{\infty} dy \int_0^h dz \left(-E_3^e q P_V + \frac{E_3^e q}{2V}P_V^2 + \beta P_0(x)q^3 P_V^3 + \frac{\beta}{4}q^4 P_V^4\right) \quad (C.5a)$$

The free energy (C.5a) **k**-representation of the trial function part $\tilde{p}(\mathbf{k},z) = P_V \cdot \tilde{q}(\mathbf{k},z)$ acquires the explicit form

$$\Delta G(P_V) = \frac{1}{S}\int_0^h dz \int_{-\infty}^{\infty} d\mathbf{k} \begin{pmatrix} -\tilde{q}\tilde{E}_3^{e*} P_V + \frac{\tilde{E}_3^{e*}\tilde{q}}{2V}P_V^2 + \frac{\beta}{4}P_V^4 \left|\int_{-\infty}^{\infty} d\mathbf{k}\tilde{q}(\mathbf{k}-\mathbf{k}',z)\tilde{q}(\mathbf{k}',z)\right|^2 + \\ \beta P_V^3 \int_{-\infty}^{\infty} d\mathbf{k}'\tilde{P}_0^*(\mathbf{k}')\tilde{q}^*(\mathbf{k}-\mathbf{k}',z)\int_{-\infty}^{\infty} d\mathbf{k}''\tilde{q}(\mathbf{k}-\mathbf{k}'',z)\tilde{q}(\mathbf{k}'',z) \end{pmatrix} \quad (C.5b)$$

Star stands for the complex conjugation.

Below we put $h \to \infty$, so external field $\tilde{E}_3^e(\mathbf{k},z) \to V \cdot \tilde{w}(\mathbf{k})k/\gamma_b \exp(-kz/\gamma_b)$ the function $\tilde{q}(\mathbf{k},z) \to \varepsilon_{33}^b \varepsilon_0 \tilde{w}(\mathbf{k})\frac{(k^2/\gamma_b^2 - s_1^2)(k^2/\gamma_b^2 - s_2^2)}{s_1 s_2 (s_1^2 - s_2^2)}(s_2 \exp(-s_1 z) - s_1 \exp(-s_2 z))$, and $\tilde{w}(\mathbf{k}) = d\frac{\exp(-kd)}{k}$. Initial polarization distribution is $\tilde{P}_0(\mathbf{k}) = P_S \frac{2\pi i L_\perp \exp(ik_1 x_0)}{\sinh(\pi k_1 L_\perp)}\delta(k_2)$.

Direct integration in Eq.(C.5b) leads to the approximate expression of free energy:

$$\Delta G(P_V) \approx \left(-P_V V + \frac{P_V^2}{2}\right)W + \left(W_3 P_V^3 + \frac{W_4}{4}P_V^4\right). \quad (C.6)$$

Where

$$W = \int_0^\infty dz \int_{-\infty}^\infty d\mathbf{k}\tilde{q}\frac{\tilde{E}_3^{e*}}{V} = 2\pi \int_0^\infty k^4 dk \frac{|\tilde{w}(\mathbf{k})|^2 \gamma_b^{-3}}{\xi s_1 s_2 (s_2 + s_1)} \frac{(s_2 + s_1 + k/\gamma_b)}{(s_1 s_2 + (s_2 + s_1)k/\gamma_b + k^2/\gamma_b^2)} \approx \pi d\sqrt{\frac{\varepsilon_0 \varepsilon_{11}}{-2\alpha_S}}$$
(C.7a)

$$W_4 = \beta \int_0^\infty dz \int_{-\infty}^\infty d\mathbf{k}\left|\int_{-\infty}^\infty d\mathbf{k}\tilde{q}(\mathbf{k}-\mathbf{k}',z)\tilde{q}(\mathbf{k}',z)\right|^2 \approx Q^2(0)\int_0^\infty dz \int_{-\infty}^\infty d\mathbf{k}|\tilde{q}(\mathbf{k},z)|^2 = 2\pi\beta Q^2(0)\times$$

$$\times \int_0^\infty k dk \varepsilon_0^2 |\tilde{w}(k)\exp(-kd)|^2 \frac{(s_1^2 + 3s_1 s_2 + s_2^2)(k^2/\gamma_b^2 - s_1^2)(k^2/\gamma_b^2 - s_2^2))^2}{2(s_1 s_2 (s_1 + s_2))^3} \approx \frac{\beta \cdot \pi d\varepsilon_{11}\varepsilon_0 \sqrt{\varepsilon_{11}\varepsilon_0}}{(-2\alpha_S)^{5/2}(L_\perp + d)^2}$$
(C.7b)

$$W_3 = \beta \int_0^\infty dz \int_{-\infty}^\infty d\mathbf{k} \int_{-\infty}^\infty d\mathbf{k}' \widetilde{P}_0^*(\mathbf{k}') \widetilde{q}^*(\mathbf{k}-\mathbf{k}',z) \int_{-\infty}^\infty d\mathbf{k}'' \widetilde{q}(\mathbf{k}-\mathbf{k}'',z) \widetilde{q}(\mathbf{k}'',z) \approx$$

$$\approx \beta Q(0) \int_{-\infty}^\infty d\mathbf{k} \widetilde{P}_0^*(\mathbf{k}) \exp(-kd) \cdot \int_0^\infty dz \int_{-\infty}^\infty d\mathbf{k}' |\widetilde{q}(\mathbf{k}',z)|^2 \approx \frac{-P_S x_0 \beta}{\sqrt{(L_\perp+d)^2 + x_0^2}} \frac{\pi d \cdot \varepsilon_{11} \varepsilon_0}{4\alpha_S^{\,2}(L_\perp+d)} \quad \text{(C.7c)}$$

In Eqs.(C.7) we substituted $Q(r) \approx \dfrac{\sqrt{\varepsilon_{11}\varepsilon_0/(-2\alpha_S)}d^2}{(L_\perp d + d^2 + r^2)\sqrt{d^2+r^2}}$. Then minimization of Eq.(C.6) on $P_V$ leads to the nonlinear equation:

$$P_V + w_3 P_V^{\,2} + w_4 P_V^{\,3} = V. \qquad (C.8)$$

Where $w_3 = 3\dfrac{W_3}{W} = \dfrac{-3\beta P_S x_0}{\sqrt{(L_\perp+d)^2 + x_0^2}} \dfrac{\sqrt{2\varepsilon_{11}\varepsilon_0}}{4(-\alpha_S)^{3/2}(L_\perp+d)}$ and $w_4 = \dfrac{W_4}{W} = \dfrac{\beta\varepsilon_{11}\varepsilon_0}{4\alpha_S^{\,2}(L_\perp+d)^2}$.

Bistability (if any) or non-zero solution of Eq.(C.8) at zero bias, $P_V(V=0) \neq 0$, should satisfy the quadratic equation $w_4 P_V^{\,2} + w_3 P_V + 1 = 0$. Real roots $P_V = \dfrac{-w_3 \pm \sqrt{w_3^2 - 4w_4}}{2w_4}$ exist under the conditions $w_3^2 \geq 4w_4$ or in evident form $\dfrac{x_0^2}{(L_\perp+d)^2 + x_0^2} \geq \dfrac{8\alpha_S}{9\alpha}$. Using expression for $\alpha_S(x_0, L_\perp) = \alpha\left(1 - \dfrac{6L_\perp(L_\perp+d)}{\pi\big((L_\perp+d)^2 + x_0^2\big)}\right)$, we obtained

$\dfrac{x_0^2}{(L_\perp+d)^2 + x_0^2} \geq \dfrac{8}{9}\left(1 - \dfrac{6L_\perp(L_\perp+d)}{\pi\big((L_\perp+d)^2 + x_0^2\big)}\right)$ that is possible at $(L_\perp+d)^2 \ll x_0^2$.

Corresponding coercive bias $V_c^\pm$ can be found from the condition $dV/dP_V = 0 = 1 + 2w_3 P_V(V_c^\pm) + 3w_4 P_V^{\,2}(V_c^\pm)$ that leads to $P_V(V_c^\pm) = \dfrac{-w_3 \pm \sqrt{w_3^2 - 3w_4}}{3w_4}$ and

$$V_c^\pm = \left(\frac{w_3 \pm \sqrt{w_3^2 - 3w_4}}{3w_4}\right)^2 \frac{\pm 2\sqrt{w_3^2 - 3w_4} - w_3}{3} = \frac{w_3(2w_3^2 - 9w_4) \pm 2(w_3^2 - 3w_4)^{3/2}}{27w_4^2}.$$

Corresponding hysteresis loop halfwidth is $\Delta V_c = \dfrac{2(w_3^2/3w_4 - 1)^{3/2}}{3(3w_4)^{1/2}}$. Using the relationship $\dfrac{w_3^2}{3w_4} = \dfrac{x_0^2}{(L_\perp + d)^2 + x_0^2}\left(\dfrac{3\alpha}{2\alpha_S}\right)$, we obtained the evident expression

$$\Delta V_c(x_0) = \dfrac{-4\alpha_S(L_\perp + d)}{3\sqrt{3\beta\varepsilon_{11}\varepsilon_0}}\left(\dfrac{x_0^2}{(L_\perp + d)^2 + x_0^2}\left(\dfrac{3\alpha}{2\alpha_S}\right) - 1\right)^{3/2}.$$

Corresponding orientation barrier existing in the range of hysteresis $V_c^- < V < V_c^+$ should be calculated from the free energy (C.5), namely $\Delta G_a = \Delta G(P_V^+, V) - \Delta G(P_V^-, V)$. For example $\Delta G(P_V, V = 0) = \dfrac{W}{24 w_4^3}\left(\pm w_3\left(w_3^2 - 4w_4\right)^{3/2} - w_3^4 + 6w_3^2 w_4 - 6w_4^2\right)$.

Obtained results may be interpreted as probe-induced domain hysteresis far from the wall, where $P_0(x) \to \mp P_S$. Actually, in dimension variables Eq.(C.8) is

$P_V - \dfrac{3\beta P_S x_0}{\sqrt{(L_\perp + d)^2 + x_0^2}}\dfrac{\sqrt{2\varepsilon_{11}\varepsilon_0} P_V^2}{4(-\alpha_S)^{3/2}(L_\perp + d)} + \dfrac{\beta\varepsilon_{11}\varepsilon_0 P_V^3}{4\alpha_S^2(L_\perp + d)^2} = V$. Introducing here the maximal induced polarization $\Delta P$ as $P_V = -2\alpha_S(L_\perp + d)\Delta P\big/\sqrt{-2\alpha_S\varepsilon_{11}\varepsilon_0}$, one can rewrite the Eq. (C.8) as follows:

$$-2\alpha_S \Delta P - \dfrac{3\beta P_S (\Delta P)^2 x_0}{\sqrt{(L_\perp + d)^2 + x_0^2}} + \beta(\Delta P)^3 = \dfrac{V}{L_\perp + d}\sqrt{\varepsilon_{11}\varepsilon_0(-2\alpha_S)}. \qquad \text{(C.9)}$$

Far from wall ($x_0 \gg d$) Eq.(C.9) simplifies as $-2\alpha\Delta P - 3\beta P_S(\Delta P)^2 + \beta(\Delta P)^3 = V\sqrt{\varepsilon_{11}\varepsilon_0(-2\alpha)}\big/(L_\perp + d)$. Then turning to full polarization $P = -P_S + \Delta P$, we obtain $\alpha P + \beta P^3 = V\sqrt{-2\alpha\varepsilon_{11}\varepsilon_0}\big/(L_\perp + d)$. Corresponding coercive biases are symmetric, namely: $V_c^\pm = \pm\dfrac{(L_\perp + d)}{\sqrt{-2\alpha\varepsilon_{11}\varepsilon_0}}\dfrac{2\alpha P_S}{3\sqrt{3}}$.

**Appendix D. Approximate analytics**

Using characteristic equation (B.2) rewritten as $\left(s^2 - \dfrac{k^2}{\gamma_b^2}\right)\left(-\dfrac{\eta k^2 - 2\alpha_S}{\xi} + s^2\right) = \dfrac{s^2}{\xi\varepsilon_0\varepsilon_{33}^b}$ and its roots properties (B.3a-b), one can rewrite Eq.(15a) as:

$$P_3(x,y,z) = P_0(x) - P_V \int_0^\infty k\,dk\, J_0\!\left(k\sqrt{x^2+y^2}\right)\widetilde{w}(k)\dfrac{(s_2\exp(-s_1 z) - s_1\exp(-s_2 z))k/\gamma_b}{\sqrt{\xi(-2\alpha_S + \eta k^2)}(s_1^2 - s_2^2)}. \quad (D.1)$$

After elementary transformations, we obtained equilibrium polarization distribution at z=0:

$$P_3(x,y,0) \approx P_0(x) + \sqrt{\varepsilon_{11}\varepsilon_0}\, P_V\, d \int_0^\infty k\,dk\, \dfrac{J_0\!\left(k\sqrt{x^2+y^2}\right)\exp(-kd)}{\sqrt{(\eta k^2 - 2\alpha_S)(1 - 2\alpha_S\varepsilon_0)}} \approx$$

$$\approx -\sqrt{\dfrac{\varepsilon_{11}\varepsilon_0}{-2\alpha_S}}\,\dfrac{P_V d^2}{(d^2+x^2+y^2)^{1/2}(d^2+x^2+y^2+L_\perp d)} \quad (D.2)$$

*Threshold bias calculations*

For the case of 1D initial profile, shown in Fig.1b, maximal local deviation from the initial profile $x_{DW}(y,z) = x_0$ appears at the sample surface $\max\{|x_0 - x_{DW}(y=0, z=0)|\} = a$, and so Eqs.(21) immediately lead to.

$$|x_{DW}(\rho) - x_0| = a = 2L_\perp\left|\mathrm{arctanh}\!\left(-\sqrt{\dfrac{\varepsilon_{11}\varepsilon_0}{-2\alpha_S(x_0,L_\perp)}}\,\dfrac{V_a d^2}{P_S(d^2+\rho^2+L_\perp d)\sqrt{d^2+\rho^2}}\right)\right|, \quad (D.3a)$$

$$y_{DW}^2(\rho) = 0 = \rho^2 - x_{DW}^2(\rho) \quad (D.3b)$$

from here $\rho^2 = x_{DW}^2(\rho)$, $x_{DW} = x_0 \pm a$ and so

$$\dfrac{a}{2L_\perp} = \left|\mathrm{arctanh}\!\left(-\sqrt{\dfrac{\varepsilon_{11}\varepsilon_0}{-2\alpha_S(x_0,L_\perp)}}\,\dfrac{V_a d^2}{P_S(d^2+(x_0\pm a)^2+L_\perp d)\sqrt{d^2+(x_0\pm a)^2}}\right)\right|, \quad (D.4)$$

Finally

$$V_{th} = \mp\sqrt{\dfrac{-2\alpha_S(x_0,L_\perp)}{\varepsilon_{11}\varepsilon_0}}\,\tanh\!\left(\dfrac{a}{2L_\perp}\right)\dfrac{P_S}{d^2}(d^2+(x_0\pm a)^2+L_\perp d)\sqrt{d^2+(x_0\pm a)^2}. \quad (D.5)$$